\documentclass[journal,twocolumn]{IEEEtran}
\usepackage{amsfonts}
\usepackage{times}
\usepackage{graphicx}
\usepackage{latexsym}
\usepackage{dsfont}
\usepackage{amssymb}
\usepackage{amsmath}
\usepackage{cite}
\usepackage{verbatim}
\usepackage{subfigure}
\usepackage{afterpage}

\def\bb0{{\mathbb{0}}}

\def\bb{{\mathbf{b}}}

\def\b0{{\mathbf{0}}}

\def\sf0{{\mathsf{0}}}

\usepackage{multirow}
\usepackage{multicol}
\usepackage{booktabs}
\usepackage{epstopdf}
\usepackage{enumerate}
\usepackage{algorithm}
\usepackage{amsmath}
\usepackage{float}
\usepackage{color}
\usepackage{makeidx}
\usepackage{bm}
\usepackage{cleveref}
\usepackage{url}
\usepackage{steinmetz}
\usepackage{varwidth}
\usepackage{soul}
\graphicspath{{figs/}}
\usepackage{tabularx}
\usepackage{gensymb}
\usepackage{balance}
\usepackage{graphicx}
\usepackage{lipsum} 
\usepackage{subcaption}
\usepackage{array}
\usepackage{booktabs}
\usepackage{url}

\makeatletter
\def\thickhline{\noalign{\ifnum0=`}\fi\hrule \@height 1.0pt \futurelet \reserved@a\@xhline}
\makeatother

\newcommand{\comm}[1]{}
\DeclareMathOperator*{\argmax}{arg\,max}

\usepackage{array} 
\begin{document}

\title{RIS-Aided mmWave O-RAN: \\  Coverage Extension and User Mobility Handling}

\author{%
    Tawfik~Osman\textsuperscript{*},~Aditya~S.~Shekhawat\textsuperscript{*},~Abhradeep~Roy,~Georgios~C.~Trichopoulos,~and~Ahmed~Alkhateeb%
    \thanks{%
        The authors are with the School of Electrical, Computer, and Energy Engineering, Arizona State University, Tempe, AZ, USA.
        Emails: \{tmosman, aditya.shekhawat, aroy59, gtrichop, alkhateeb\}@asu.edu.
        This work was supported by the National Science Foundation (NSF) Grant No. 2229530.\\
        \textsuperscript{*}These authors contributed equally to this work.%
    }%
}

\maketitle

\begin{abstract}
Reconfigurable Intelligent Surfaces (RISs) can redirect electromagnetic waves to desired directions to enhance signal coverage and/or improve the signal-to-noise ratio (SNR) at the user equipment (UE). 
We present the design, implementation, and evaluation of an RIS-assisted O-RAN 5G system operating in the FR2 millimeter-wave (mmWave) frequency band. 
We first introduce the design of a 1{,}024-element ($32 \times 32$), 1-bit RIS operating at the 28~GHz band, utilizing a modular and scalable tiled architecture. 
Then, we demonstrate how the O-RAN E2 interface can be leveraged to dynamically control RIS configurations without modifying standard 5G signaling procedures. 
To evaluate the RIS-assisted 5G system, we conducted extensive field trials in both indoor and outdoor environments. 
The results of the O-RAN link coverage trials show that the deployed RIS provides substantial received signal power gains, ranging from 9--20~dB and 6--18~dB in indoor and outdoor scenarios, respectively. 
Handling UE mobility in RIS-assisted systems is challenging due to the need for joint RIS and UE beam management. 
For that, we develop two UE mobility management algorithms and evaluate them in real-time operation using the RIS O-RAN testbed. 
These algorithms leverage the received signal power at the UE to jointly track and adapt the RIS and UE beams in real time as the UE moves. 
The findings draw important insights into the practical feasibility of integrating RIS into O-RAN systems to enhance coverage, mobility support, and link reliability in next-generation cellular networks.
\end{abstract}


\begin{figure*}[t]
    \centering
    \subfigure[]{
        \includegraphics[width=0.4\textwidth]{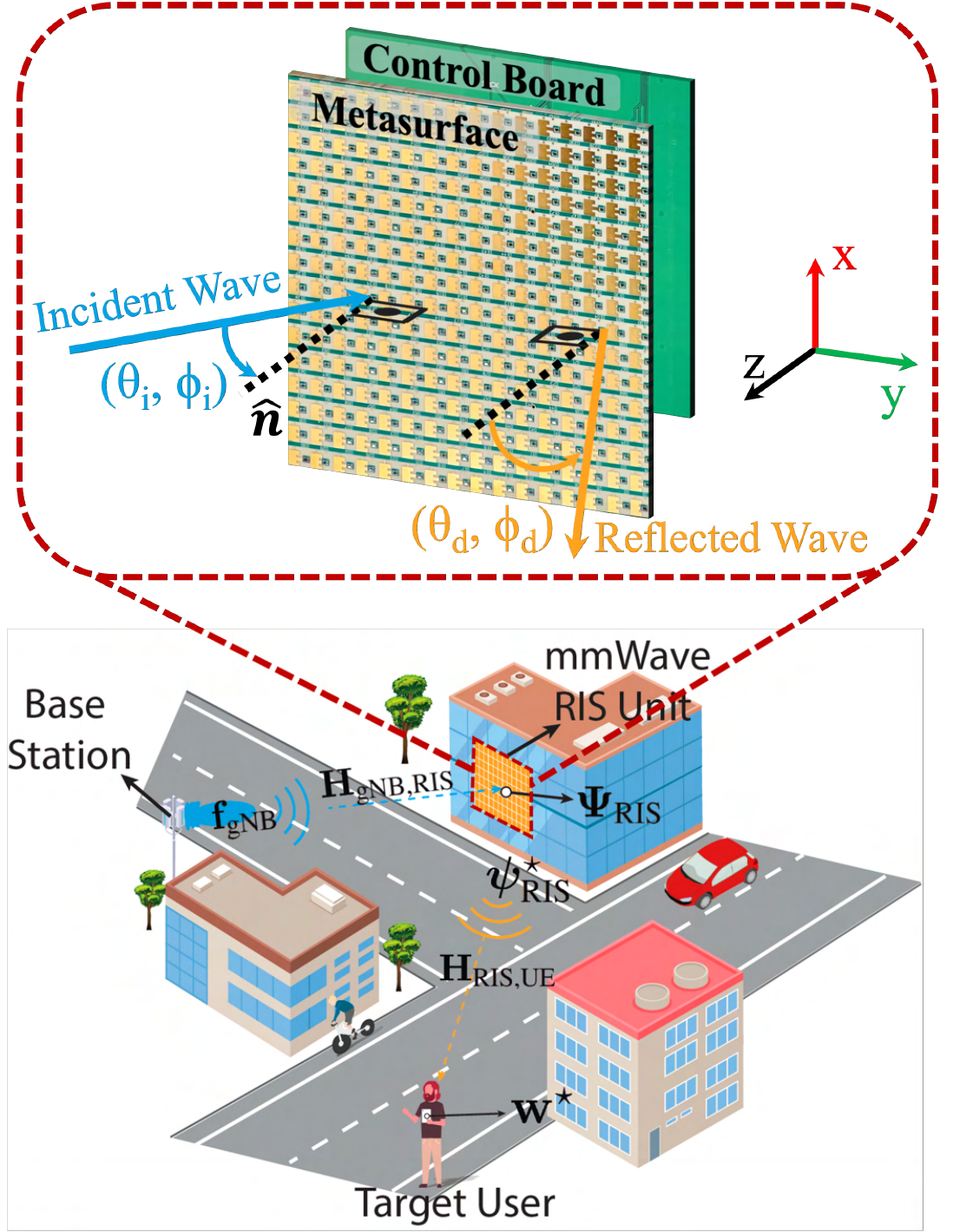}
        \label{fig:system_model}
    }
    \subfigure[]{
        \includegraphics[width=0.55\textwidth]{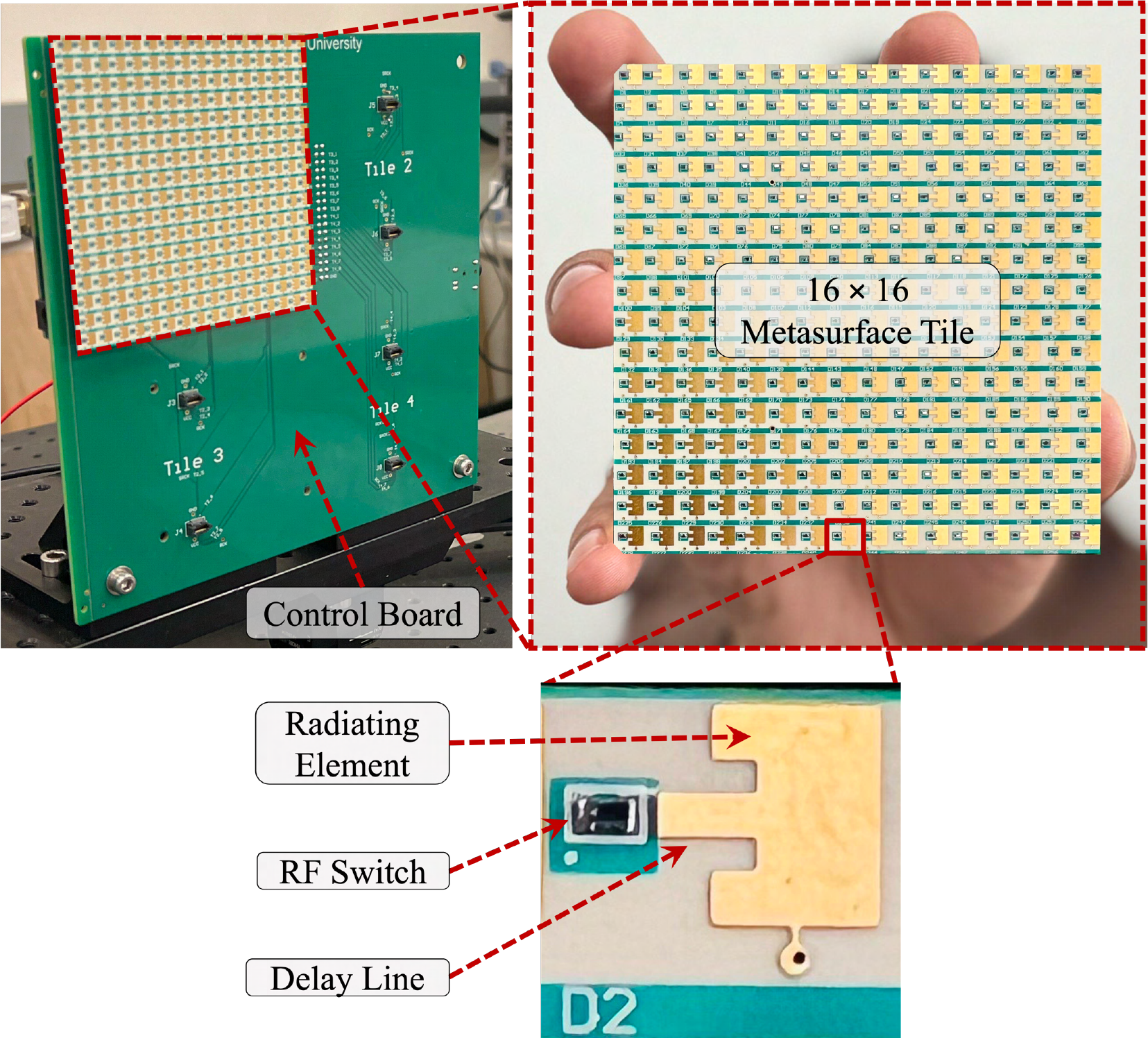}
        \label{fig:mmWRIS_prototype}
    }
    \caption{(a) This figure depicts the system model for an RIS-assisted FR2 (mmWave) 5G SA communication system. The base station, operating in time-division duplex (TDD) mode, communicates with a target user in the presence of tall buildings obstructing the direct link. The RIS is strategically deployed to establish a virtual line-of-sight (LoS) link, enhancing signal strength and coverage in the blocked region. (b) The RIS prototype used in this work consists of 4 tiles, each hosting 256 radiating element ($16 \times 16$). The tiles are attached to a control board which facilitates switch biasing and the integration of the RIS into the ORAN system.}
    \label{fig:figure1}
\end{figure*}

\section{Introduction} \label{sec:Intro}
The deployment of fifth-generation (5G) wireless networks enabled higher data rates, lower latency, and massive device connectivity through the adoption of new technologies, such as directional communication over millimeter wave (mmWave) frequency bands using large antenna arrays \cite{3gpp2018nr,Heath2016,Alkhateeb2014e}. However, while mmWave communication offers these new performance gains, its deployments face new challenges with coverage gaps and dead zones caused by obstacles and unfavorable propagation conditions  \cite{rappaport2013mmwave, molisch2017mmwave}. Therefore, beyond 5G and 6G systems need to develop new approaches and technologies to overcome these challenges. A promising solution towards this goal is the deployment of reconfigurable intelligent surfaces (RISs) \cite{wu2020irs,direnzo2019sre}, which can reconfigure the wireless propagation environment by intelligently controlling the reflection of electromagnetic waves and redirecting them towards desired directions, which is the motivation of this work.

\subsection{Brief Background on RIS} \label{subsec:Brief Background}
RISs comprise two main components: the metasurface and the corresponding control circuitry. The metasurface consists of densely arranged, tunable subwavelength radiating structures, such as microstrip patch antennas that control the phase and/or amplitude characteristics of reflected electromagnetic waves. Unlike conventional smooth reflectors (e.g., mirrors) that exhibit specular reflection due to uniform phase delays as the wave traverses the surface, RISs enable anomalous reflection, where the reflection angle can differ from the incident angle. This capability is achieved by dynamically modifying the phase and/or amplitude properties of the reflected wavefront \cite{yu_2011}. Each subwavelength unit cell within the metasurface captures incoming signals and re-radiates them with a tailored phase and/or amplitude profile. This wavefront modulation is typically realized by integrating active components, such as radio-frequency (RF) switches (e.g., PIN diodes, varactors, and MEMs) within the radiating elements. These switches can alter the path of electric current within each cell (open/short termination), thereby controlling the phase and/or amplitude characteristics of the re-radiated signals. Incorporating multiple switches per cell can facilitate multi-bit modulation, further enhancing beam steering precision and overall efficiency. Beyond simply redirecting signals to desired directions in the far-field, RISs can also achieve signal focusing when the user equipment (UE) or base station (BS) is positioned within the radiating near-field region of the RIS. 

Several RIS architectures have been reported in the literature, featuring diverse configurations that may involve single or multiple active elements, as well as single-layered or multilayered substrates \cite{hum_2014}. Regardless of their specific configuration, all RIS implementations require dedicated control circuitry composed of a biasing circuit and a control unit (such as a microcontroller) to precisely manage the operational states of the tunable elements via adjusting the biasing voltage across the device terminals. Structurally, RISs are considered two-dimensional surfaces because the lateral dimensions span several wavelengths, while their thickness typically remains a fraction of the wavelength. These geometrical attributes render RISs ideal for integration onto various surfaces, including indoor and outdoor installations, as well as deployment on curved building facades.

At mmWave and THz frequencies, signal propagation in non-line-of-sight (NLoS) scenarios relies heavily on diffuse scattering from rough surfaces, such as walls or terrain, especially when the user is located away from the specular reflection direction \cite{ma_2019,cui_2020}. Under these conditions, the received signal power at the UE typically experiences significant degradation. To mitigate this issue, RISs can exploit anomalous reflection to create alternative propagation paths, thereby enhancing the received signal strength. The received power \( P_r \) when a RIS is strategically placed between the BS and the UE (as illustrated in Fig.~\ref{fig:figure1}) can be estimated using the bistatic radar equation:

\begin{equation}\label{eq:received_power}
P_r = \frac{P_t G_{BS} G_{UE} \lambda^2 \sigma}{(4\pi)^3 R_i^2 R_d^2}
\end{equation}

where $P_t$ denotes the transmit power of the BS, and $G_{\text{BS}}$ and $G_{\text{UE}}$ represent the antenna gains of the BS and the UE, respectively. The distances between the BS and RIS and between the RIS and UE are defined as $R_i$ and $R_d$, respectively. The radar cross-section (RCS) $\sigma$ of RIS can be approximated as that of a flat rectangular conductive surface with losses. For an electrically large RIS with an area $A$, efficiency $\eta$, and wavelength $\lambda$, the maximum monostatic RCS is:

\begin{equation}\label{eq:monostatic_rcs}
\sigma = \frac{4 \pi \eta A^2}{\lambda^2}
\end{equation}

In practical wireless communication scenarios, which typically involve bistatic configurations, the RIS is illuminated by the BS at an incident angle $(\theta_i, \phi_i)$ and observed from the UE at a reflection angle $(\theta_d, \phi_d)$ (see Fig.~\ref{fig:figure1}). The bistatic RCS of the RIS can be approximated as follows:

\begin{equation}\label{eq:bistatic_rcs}
\sigma = \frac{4 \pi \eta A^2 (\hat{n} \cdot \hat{k}_i)(\hat{n} \cdot \hat{k}_d)}{\lambda^2}
\end{equation}

where, $\hat{k}_i$ and $\hat{k}_d$ denote the vectors specifying the directions of propagation of the incident and reflected waves, respectively. The $\hat{n}$ represents the unit normal vector to the RIS aperture. For the configuration illustrated in Fig.~\ref{fig:figure1}a, these vectors can be expressed as:

\begin{subequations}\label{eq:unit_vectors}
\begin{align}
\hat{k}_i &= (\sin\theta_i \cos\phi_i,\ \sin\theta_i \sin\phi_i,\ \cos\theta_i) \label{eq:k_i_vector} \\
\hat{k}_d &= (\sin\theta_d \cos\phi_d,\ \sin\theta_d \sin\phi_d,\ \cos\theta_d) \label{eq:k_d_vector} \\
\hat{n}   &= (0,\ 0,\ 1) \label{eq:n_vector}
\end{align}
\end{subequations}

Therefore, the bistatic RCS of the RIS can be simplified to:

\begin{equation}\label{eq:bistatic_rcs}
\sigma = \frac{4 \pi \eta \cos\theta_i \cos\theta_d A^2}{\lambda^2}
\end{equation}

It can be observed that the received power increases quadratically with the size of the RIS; specifically, enlarging the RIS area (A) by a factor of ten provides a 20 dB improvement in received signal strength under identical propagation conditions. Additionally, the received signal power is inversely proportional to the square of the operating wavelength, indicating significant advantages when employing RISs at higher frequencies. This property becomes particularly important as signals at higher frequencies undergo greater attenuation, primarily due to increased free-space path loss and higher penetration losses through common building materials. For example, at 38 GHz, the attenuation through window glass and brick walls can exceed 25 dB and 91 dB, respectively \cite{rodriguez_2015}.

\subsection{Prior Work} \label{subsec:Prior Work}

In recent years, the design and analysis of wireless communication systems incorporating RISs have attracted significant research attention. From a signal-processing perspective, RIS-based systems introduce unique challenges, particularly related to designing large-dimensional passive beamforming matrices for controlling reflected signals. Several recent studies have addressed these challenges: for instance, \cite{huang_2019} focuses on developing low-complexity and energy-efficient RIS beamforming algorithms; \cite{zhou_2021} investigates robust beamforming solutions; and \cite{ye_2020,zhu_2021,yue_2021} explore solutions for jointly optimizing beamforming at both the base station and RIS. However, a major operational challenge associated with these surfaces is the considerable training overhead required for accurate channel estimation, particularly because RISs are predominantly passive devices. To mitigate this issue, RIS designs incorporating sparsely distributed active elements have been proposed in \cite{taha_2021}, enabling partial channel state acquisition. These architectures can further exploit compressive sensing and deep learning methods for efficient RIS channel estimation.

From the network-level viewpoint, previous research has examined the impact of RIS deployment on coverage performance in various scenarios, including pure reflective configurations for indoor and outdoor environments \cite{ying_2020,nemati_2020_coverage,moro_2020,he_2021}, refraction-based approaches to improve outdoor-to-indoor propagation \cite{nemati_2021_modeling}, and combined transmission and reflection schemes \cite{mu_2021_star,xu_2021_star}. Nevertheless, most of these prior investigations \cite{taha_2021,ying_2020,nemati_2020_coverage,moro_2020,he_2021,nemati_2021_modeling,huang_2019,zhou_2021,ye_2020,zhu_2021,yue_2021} primarily rely on numerical simulations rather than experimental validation. Therefore, to rigorously quantify the practical benefits offered by RIS technology, it is essential to develop proof-of-concept prototypes and conduct real-world experiments that measure coverage enhancement and achievable data-rate improvements in realistic wireless environments.

From the perspective of practical realization and experimental validation, research on RIS implementations, remains relatively limited. Existing studies have investigated RIS designs at sub-6 GHz \cite{pei_2021, george_2022, wu_2022, liang_2022}, around 10 GHz \cite{zhang_2022, cao_2023, yang_2016, cui_2014, wan_2016} and at mmWave frequencies \cite{gros_2021, wang_2024, shekhawat_2022, shekhawat_aps_2022, jeong_2022, tang_2022, oh_2023, shekhawat_ims_2024, shekhawat_2024, shekhawat_2025, kamoda_2011}, often employing PIN and varactor diode based architectures. Additionally, interest is steadily growing in extending RIS technology for THz frequencies \cite{headland_2017}. A common limitation faced by current implementations is scalability. Existing RIS designs are often constrained to one-dimensional (1D) electronic beam steering configurations, as highlighted in \cite{pei_2021, liang_2022, zhang_2022}, or are limited in aperture size by the practical constraints on the number of biasing lines that can be incorporated into 2D planar surfaces. Increasing the biasing complexity generally necessitates additional printed circuit board (PCB) layers, thereby driving up fabrication costs and reducing practical feasibility. For RIS technology to be truly effective in wireless communication applications, designs must be scalable to large surfaces encompassing thousands of unit cells and switches. Additionally, these surfaces must maintain a low-profile and conformal geometry suitable for integration on indoor and outdoor structures.

Another key limitation associated with RISs employing quantized coding schemes is the emergence of a grating/quantization lobe, particularly evident under plane-wave illumination. This lobe results from periodicity in the phase rounding error, which is inherent in low-bit quantization schemes, introducing an undesired secondary beam that degrades overall system performance \cite{smith_1983}. The quantization lobe is especially problematic in 1-bit quantization schemes, where its power is comparable to the main beam \cite{liu_2017, yang_2016, alnuaimi_2018} and appears symmetrically opposite the specular reflection angle. This phenomenon complicates practical deployment due to increased risks of signal leakage, interference, and susceptibility to eavesdropping, as highlighted in \cite{zhao_2024}. Therefore, mitigating the quantization lobe is crucial to minimizing unwanted signal dispersion, ensuring that energy remains concentrated in the intended direction.

To address these challenges, we have developed an RIS topology that mitigates the quantization lobe in planar 1-bit RIS while providing a modular, tiled RIS architecture that is highly scalable and cost-effective. Specifically, we utilize the concept of random pre-phasing, originally developed for multi-bit phased array antennas \cite{pei_2021, shekhawat_2025}, to disrupt the periodicity in phase rounding errors. This is achieved by embedding pseudo-random pre-coded phase delays within each unit cell, effectively reducing quantization lobes. Furthermore, achieving scalability in RIS implementations requires careful consideration of the control circuitry design. As previously noted, many existing control architectures either impose constraints on the array size due to limitations in the number of biasing lines that can be integrated into the aperture or exhibit inefficiencies in power consumption. This underscores the need for a control topology that is both scalable and power-efficient while remaining practical for fabrication, particularly for managing phase shifters (e.g., PIN diodes) in single- and multi-bit RIS configurations. To address these requirements, we designed an optimized control topology that employs low-power series-in and parallel-out shift register (SR) integrated circuits (ICs) within the RIS. Furthermore, the proposed topology is inherently adaptable to higher-frequency and multi-bit RIS architectures, offering a versatile and scalable solution for next-generation wireless communication applications.

To evaluate the RIS design under realistic system conditions, we integrated our RIS system into an end-to-end cellular communication system. This integration aims to demonstrate the practical deployment benefits of our RIS, particularly in coverage extension and UE mobility support over RIS-assisted links. To enable this evaluation within a flexible and programmable environment, we adopt the open radio access network (O-RAN) architecture, which integrates a modular and disaggregated RAN design with standardized interfaces into the existing 5G stack. O-RAN also facilitates multi-vendor interoperability and supports both external and AI-driven control through RAN Intelligent Controllers (RICs).\cite{polese2022understandingoranarchitectureinterfaces}. These features make O-RAN a promising architectural framework for integrating and evaluating RIS-assisted functionalities in real-world 5G systems.

The O-RAN framework is developed and maintained by the O-RAN Alliance, and is widely adopted in the deployment of 5G systems that comply with 3GPP standards. The key elements of the O-RAN framework include the Service Management and Orchestration (SMO) layer, the near-real-time RAN Intelligent Controller (near-RT RIC), and standardized interfaces such as E2 and A1~\cite{oran-wg1}. These components enable real-time network monitoring, policy-driven optimization, and the deployment of xApps—software-defined applications that implement data-driven functions such as traffic steering, energy savings, and radio resource management~\cite{Hoffmann_2023}. Leveraging these standardized interfaces, recent research has explored the integration of external devices into O-RAN-compliant systems for intelligent control and monitoring. For example, the work in~\cite{Anapana_2024} proposed and validated the use of an E2-like interface to control a standalone mmWave phased array module, enabling 5G waveform transmission at mmWave frequencies and near-real-time beamforming adaptation. Similarly,~\cite{sahin2025rismeetsoranpractical} demonstrated the integration of a sub-6~GHz RIS into a 5G system using E2 interface feedback to optimize reflection control based on real-time RAN measurements.

Despite recent advances, prior work on real-world mmWave O-RAN testbeds tailored for research has not incorporated RIS into the end-to-end system deployment. Furthermore, existing practical implementations of RIS-assisted O-RAN are primarily limited to sub-6~GHz RIS configurations. Extending O-RAN operation to the FR2 band with integrated RIS introduces additional system design and hardware integration challenges, including joint RIS and UE beam tracking, low-latency control, and synchronization constraints across RF front-end modules. These gaps motivate the contributions of this paper, which focus on the design, implementation, and evaluation of a mmWave O-RAN system integrated with RIS under realistic deployment conditions.

\subsection{Contribution} \label{subsec:contribution}

This paper presents empirical research conducted in diverse environments using a real-time programmable RIS system with 1-bit phase resolution, integrated into an O-RAN-compliant 5G system operating in the FR2 band. The experiments evaluate the performance of the RIS in both indoor and outdoor scenarios, addressing key challenges related to coverage extension, mobility support and practical system integration. The main contributions of this paper are highlighted as follows:

\begin{itemize}
    \item \textbf{mmWave RIS Prototype:} We present a multi-layer, 1-bit RIS prototype operating at mmWave frequencies (27.2 GHz). The architecture features a modular, tile-based design optimized for scalable fabrication. It demonstrates reliable electronic beam-steering performance in both azimuth and elevation planes, incorporating randomized pre-phases onto each unit cell to effectively mitigate quantization/grating lobes.
   
    \item \textbf{End-to-End mmWave RIS-Aided Communication System:} We establish an end-to-end RIS-assisted communication system operating in the FR2 band, with near real-time RIS control enabled through standardized O-RAN interfaces. The system collects periodic reference signal received power (RSRP) measurements via the E2 interface and utilizes this information to adjust RIS beam configurations accordingly. This approach enables effective RIS beam steering toward intended users, ensuring robust connectivity in coverage-challenged scenarios.

    \item \textbf{Field Trials for Evaluating RIS Deployment in Non-Line-of-Sight Scenarios:} We present results from extensive field trials conducted to evaluate the performance of the proposed RIS-assisted O-RAN framework in coverage-challenged environments. The trials included both indoor and outdoor scenarios, where the user followed predefined trajectories while link performance metrics were continuously collected and subsequently post-processed. The results validate the significance of the deployed RIS system, in enhancing link reliability and signal quality, and provide practical insights for the development of standard-compliant RIS control and management algorithms.

    \item \textbf{Mobility Management and Connectivity Support for Mobile UE:} We evaluate two RIS beam management approaches aimed at sustaining continuous connectivity and stable throughput for mobile users. Both approaches leverage received signal power measurements and were validated through experimental trials. Their respective advantages and limitations were analyzed to guide the design of future UE mobility-aware RIS control algorithms.

\end{itemize}

The remainder of this paper is organized as follows. Section~\ref{sec:ris-aided_comm} presents the adopted system model for the base station (BS) and user equipment (UE), as well as the quantized beamforming theory of RIS and its relationship to BS--UE wireless communication. Section~\ref{sec: RIS DESIGN} describes the design and characterization of the mmWave RIS, including the structure of its building blocks, the characterization of RIS beamforming codewords, and the integration with the required control circuitry. In Section~\ref{sec:proposed_soln}, we discuss the motivation for adopting RIS-assisted communication within the O-RAN architecture and detail the implementation of the proposed mmWave RIS--O-RAN system. Section~\ref{sec:field} introduces the field measurement campaigns that demonstrate the impact of RIS on coverage enhancement. Section~\ref{sec:mobility} addresses the UE mobility tracking challenge and presents proposed solutions based on received signal strength. Section~\ref{sec:results} reports the results of the field trials and the performance of the mobility management algorithms. Finally, Section~\ref{sec:conc} concludes the paper and outlines directions for future research.

\textbf{Notation:} \( \mathbf{A} \) is a matrix, \( \mathbf{a} \) is a vector, \( a \) is a scalar, and \( \mathcal{A} \)  is a set of vectors. \( \operatorname{diag}(\mathbf{a}) \) is a diagonal matrix with entries of \( \mathbf{a} \) on its diagonal. \( |\mathbf{A}| \) is the determinant of \( \mathbf{A} \), \( \mathbf{A}^T \) is its transpose, \( \mathbf{A}^H \) is its Hermitian. \( \mathcal{N}(\mathbf{m}, \mathbf{R}) \) is a complex Gaussian random vector with mean \( \mathbf{m} \) and covariance \( \mathbf{R} \).



\section{RIS-Aided Wireless Communication System}\label{sec:ris-aided_comm}

\subsection{System and Signal Models}\label{sec:sys_ch_mod}

This section describes the communication scenario illustrated in Fig.~\ref{fig:system_model}, where a mmWave MIMO base station (gNB) communicates with a mmWave MIMO UE via a RIS. The RIS is modeled as a two-dimensional array consisting of \( N \times N \) reflecting elements, where  \( N \) represents the number of radiating RF elements along each dimension. Both the gNB and UE operate in 5G Standalone (SA) mode, where the communication infrastructure is fully reliant on the 5G Core (5GC) without dependency on legacy LTE networks. Additionally, the communication follows a time division duplex (TDD) scheme, where the same frequency resources are shared between uplink and downlink transmissions but separated and synchronized in time.

Considering the downlink transmission, we modeled the transmitted signal at the gNB, \( \mathbf{s} \) as a complex signal vector,  \( \mathbf{s} \in \mathbb{C}^{ N_\mathrm{SC} \times 1} \) and the signal power satisfies the constrained equation in \eqref{eq:power_constraint}\:
\begin{equation}
\mathbb{E} \left[ \|\mathbf{s}_t\|_2^2 \right] \leq P_t, \, \forall t \in \{1, \dots, M_{\text{gNB}}\},
\label{eq:power_constraint}
\end{equation}
where \( N_\mathrm{SC} \) represents the number of subcarriers, \( \mathbf{s}_t \) and \( P_t \) denote the complex signal vector  and the power limit for each transmit antenna respectively,  and \( M_{\text{gNB}} \) is the number of antenna elements at the gNB. The received signal at the UE, denoted by \( y \), is expressed in \eqref{eq:received_signal}:
\begin{equation}
y = \mathbf{w}_{\text{UE}}^H \big( 
\underbrace{\mathbf{H}_{\text{gNB,UE}} \mathbf{f}_{\text{gNB}} \mathbf{s}}_{\text{Direct/Multipath Link}} 
+ 
\underbrace{\mathbf{H}_{\text{RIS,UE}} \mathbf{\Psi}_{\text{RIS}} \mathbf{H}_{\text{gNB,RIS}} \mathbf{f}_{\text{gNB}} \mathbf{s}}_{\text{RIS-Assisted Link}}
\big) 
+ \mathbf{n} ,
\label{eq:received_signal}
\end{equation}
where  \( \mathbf{s}\) denotes the transmitted signal, \( \mathbf{f}_{\text{gNB}}  \in \mathbb{C}^{ M_{\text{gNB}} \times 1} \) represents  the transmit beamforming vector at the gNB,  \( \mathbf{H}_{\text{gNB,UE}} \in \mathbb{C}^{ M_{\text{UE}} \times M_{\text{gNB}}} \) represents the direct channel gain matrix between the gNB and the UE, with \( M_{\text{UE}} \) denoting the number of antenna elements UE. The matrices \( \mathbf{H}_{\text{gNB,RIS}} \in \mathbb{C}^{ N^2 \times M_{\text{gNB}} } \) and \( \mathbf{H}_{\text{RIS,UE}} \in \mathbb{C}^{ M_{\text{UE}} \times N^2 } \) denote the gNB-to-RIS and RIS-to-UE channel matrices, respectively. The receive combining vector at the UE is \( \mathbf{w}_{\text{UE}}  \in \mathbb{C}^{1  \times M_{\text{UE}}} \). The RIS phase configuration matrix is denoted by \( \mathbf{\Psi}_{\text{RIS}}  \in \mathbb{C}^{ N^2 \times N^2}  \) , where \( \mathbf{\Psi}_{\text{RIS}} = \operatorname{diag}(\boldsymbol{\psi}_{\text{RIS}}) \), and \( \boldsymbol{\psi}_{\text{RIS}} \in \mathbb{C}^{ N^2 \times 1} \) is the RIS interaction vector. Finally, we modeled the UE receiver hardware noise level, \(\mathbf{n}\) as Additive white Gaussian noise is represented by \( \mathbf{n} \sim \mathcal{CN}(0, \sigma^2) \).


\subsection{1-Bit Beamforming Codebook with Random Pre-Phasing Technique} \label{sec:1-BIT BEAMFORMING CODEBOOK}
Let us examine an $N \times N$ array of radiating elements arranged in a two-dimensional planar structure on the $x$-$y$ plane. In many practical deployments, an RIS is typically positioned far from the transmitter, such as a base station, and is illuminated by a plane wave. To control the redirection of an incoming plane wave arriving from $(\theta_i, \phi_i)$ toward a target direction $(\theta_d, \phi_d)$, the phase shift $\phi_{mn}$ required on every $mn^{\text{th}}$ element can be expressed as:

\begin{equation}\label{eq:phase_shift}
\phi_{mn} = \phi_{d,mn} - \phi_{i,mn}
\end{equation}

where $\phi_{d,mn}$ denotes the progressive phase required to steer the main beam toward the desired reflection direction, and $\phi_{i,mn}$ represents the phase shift induced by the incident wave at the $mn^{\text{th}}$ element. These phases can be expressed as:

\begin{align}
\phi_{d,mn} &= k_\circ \left( x_m \sin\theta_d \cos\phi_d + y_n \sin\theta_d \sin\phi_d \right) \label{eq:progressive_phase_shift} \\
\phi_{i,mn} &= k_\circ \left( x_m \sin\theta_i \cos\phi_i + y_n \sin\theta_i \sin\phi_i \right) \label{eq:illumination_phase_shift}
\end{align}

where $k_\circ$ represents the free-space wavenumber, and $(x_m, y_n)$ denote the coordinates of the $mn^{\text{th}}$ element within the array. At mmWave and THz frequencies, practical phase-shifting architectures are limited to generating discrete phase values, which are typically quantized using 1-bit, 2-bit, or 3-bit schemes. In this work, a single-bit phase quantization approach is employed due to its simplicity and cost-effectiveness compared to higher-bit quantization methods \cite{yang_2017}.

For a 1-bit quantization scheme, the phase values of $\phi_{mn}$ are discretized to either $0^\circ$ or $180^\circ$. This, however, leads to periodic phase errors and appearance of undesired quantization lobes. To mitigate this issue, random phase variations, denoted as $\phi_{mn}^{\text{rand}}$, are introduced at each element through delay lines \cite{shekhawat_2025}. Although these phase shifts are fixed and cannot be modified post-fabrication, they effectively suppress quantization lobes and ensure that the RIS primarily reflects energy in the desired direction. With these modifications, the continuous excitation phase in ~\eqref{eq:phase_shift} can be redefined as:

\begin{equation}\label{eq:modified_phase}
\phi_{mn} = \phi_{d,mn} - \phi_{i,mn} - \phi_{rand,mn}
\end{equation}

Here, $\phi_{rand,mn}$ represents an $N \times N$ matrix containing random phase delays generated using a uniform pseudo-random number generator, expressed as:

\begin{equation}\label{eq:random_phase}
\phi_{rand,mn} = \text{rand}(m,n) \cdot 180^\circ
\end{equation}

Following this, the continuous excitation phase is discretized to either $0^\circ$ or $180^\circ$, given by:

\begin{equation}\label{eq:quantized_phase}
\phi_{mn}^{\text{quant}} = 180^\circ \cdot \text{floor} \left( \frac{\phi_{mn}}{180^\circ} + 0.5 \right)
\end{equation}

For elements in the OFF state, all phase values within the range $[-90^\circ, 90^\circ]$ are mapped to $0^\circ$ (state 1), whereas values outside this range are assigned to $180^\circ$ (state 2) for the ON state. A detailed explanation of how these states are implementation on RIS using PIN diode is provided in section \ref{sec: RIS DESIGN}. The resulting radiation pattern is determined by analyzing the array factor of the RIS, which is represented as:

\begin{equation}\label{eq:array_factor}
AF_{\text{RIS}} (\theta,\phi) = \sum_{i=1}^{M} \sum_{j=1}^{N} e^{-j\phi_{mn}^{\text{total}}} e^{-jk_0 (x_m u + y_n v)}
\end{equation}

where $ \phi_{mn}^{\text{total}} = \phi_{mn}^{\text{quant}} + \phi_{i,mn} + \phi_{rand,mn}$ is the total phase at each element and the term $k_0 (x_m u + y_n v)$ represents the phase modulation introduced by Green’s function, where $u = \sin\theta \cos\phi$ and $v = \sin\theta \sin\phi$.


For the random phases $\phi_{rand,mn}$, a $N \times N$ random delay matrix is generated, where each element takes a value between $0^\circ$ and $180^\circ$. However, to ensure these randomly assigned delays effectively minimize quantization lobes across different incident and reflected wave angles, the side-lobe level (SLL) in the radiation pattern is analyzed to determine the optimal $\phi_{rand,mn}$ \cite{shekhawat_2025}. The far-field radiation characteristics of the RIS are computed using the array factor presented in \eqref{eq:array_factor}.



\begin{figure}
\centerline{\includegraphics[width=3.5in]{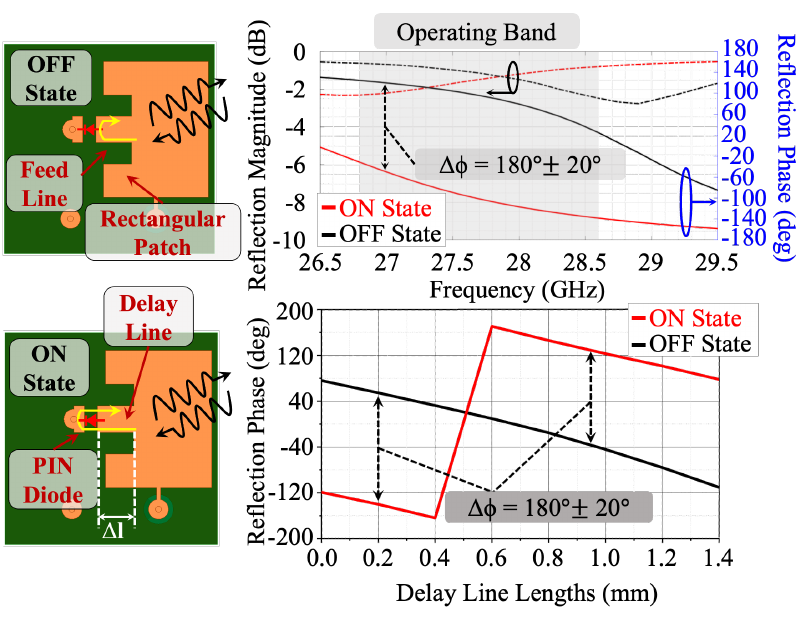}}
    \caption{The current distribution during ON and OFF state (left), resulting in phase modulation of the reflected signal ($180^\circ$). Reflection magnitude and phase response of the unit cell for ON and OFF states (top right). The phase difference remains $180^\circ \pm 20^\circ$ within a 1.8 GHz bandwidth. Phase variation at 27.7 GHz for different delay line lengths in the ON/OFF states (bottom right).}
    \label{fig:unit_cell_and_results}
\end{figure}

\section{mmWave RIS Design} \label{sec: RIS DESIGN}

The proposed RIS design consists of a multi-layered architecture involving careful engineering of the element, metasurface, and associated control circuitry to achieve optimal beamforming performance. Each of these components plays a critical role in precisely manipulating the phase of the incident electromagnetic wave for dynamic beam steering. In the subsequent subsections, we present detailed discussions on the design considerations and practical implementation for each of these key components.

\subsection{Element, Metasurface, and Control Board Design}

The unit cell forms the fundamental element of the RIS, allowing accurate control of phase of the incident electromagnetic waves to achieve specific reflection properties. The unit cell employs four metallization layers to incorporate all the required components and has a dimension of \( \lambda_0/2 \times \lambda_0/2 \), with \(\lambda_0\) representing the free-space wavelength. A full-wave electromagnetic simulation was performed using ANSYS Electronics Desktop (HFSS) to evaluate the reflection magnitude and phase characteristics of the element. The dimensions of both the radiating element and its feed line were optimized to achieve a targeted phase difference of $180^\circ \pm 20^\circ$ while ensuring magnitude variations remained below 1 dB, as illustrated in Fig.~\ref{fig:unit_cell_and_results}. The diode (MACOM MA4AGP907) switching states (ON/OFF) were incorporated into the simulation model using measured S-parameters provided by the vendor.

Furthermore, to introduce phase randomization and effectively suppress quantization lobes, pseudo-random pre-phase delays from $0^\circ$ to $180^\circ$ are generated using equation \eqref{eq:random_phase}, which are implemented onto each unit cell by adjusting the inset feed length \(\Delta l\) (as shown in Fig. \ref{fig:unit_cell_and_results}). The results presented in Fig.~\ref{fig:unit_cell_and_results} (bottom right) are normalized relative to the shortest delay line length. By varying the delay line length ($\Delta l$) from 0 mm to 1.4 mm, the desired range of random phase delays can be precisely achieved. Additionally, as depicted in Fig.~\ref{fig:unit_cell_and_results}, this approach consistently preserves the required phase difference of $180^\circ \pm 20^\circ$ between the diode states.

\begin{figure}
    \centering
    
    \subfigure[]{ \includegraphics[width=0.3\textwidth]{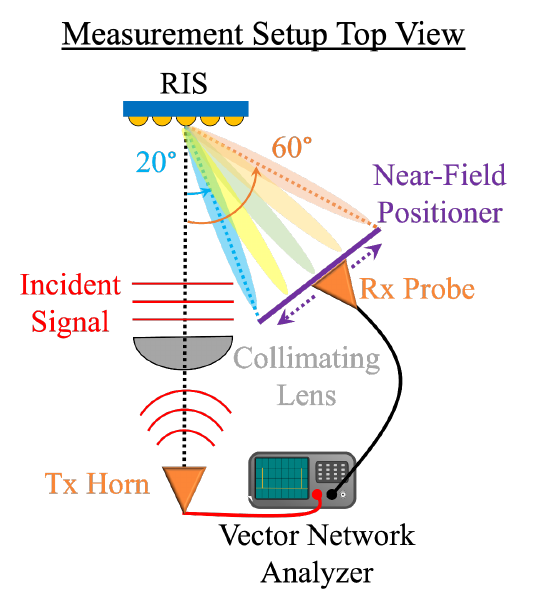}\label{fig:mmWRIS_measurement_setup}} 
    
    \subfigure[]{ \includegraphics[width=0.5\textwidth]{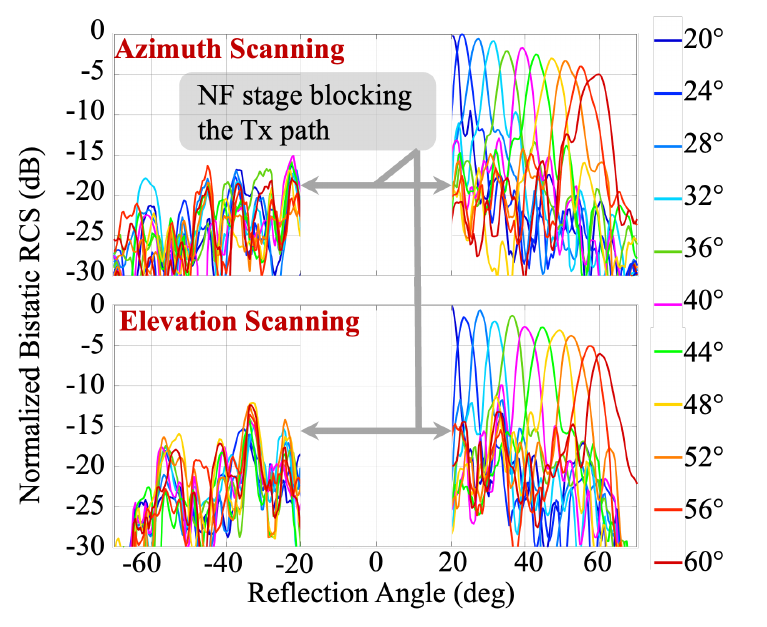} \label{fig:mmWRIS_measurement_results}}
    
    \caption{(a) Beam characterization setup and (b) measured reflected beams in the azimuth and elevation planes.}
    
    \label{fig:mmWRIS_beam_characterization} 
\end{figure}

Following the evaluation of the element design, we next analyze its performance within a finite array configuration. A \(16\times16\) metasurface is designed by periodically arranging unit cells, resulting in a rectangular lattice with a spacing of \(\lambda_0 = 5.4\) mm per tile. The tile based structure ensures scalability while maintaining a compact, multi-layer PCB configuration, which minimizes potential warpage issues such as bow or twist during fabrication. For the initial prototype, a \(32\times32\) RIS array comprising 1,024 elements is constructed from four individual RIS tiles. Nevertheless, the modular design allows the integration of an arbitrary number of tiles without imposing additional fabrication or integration constraints.

To introduce randomized pre-phases, an optimized $32\times32$ matrix of random values was generated according to equation~\eqref{eq:random_phase} and implemented on every unit cell by adjusting the delay line lengths. Details, including the respective random pre-phases for each tile and full-wave electromagnetic simulations to analyze the bistatic RCS to evaluate the beam steering performance and quantization lobe suppression, have been thoroughly conducted and are comprehensively documented in \cite{shekhawat_2025}. Following the numerical analysis of the $16\times16$ finite arrays based on the proposed element design, the four RIS tiles were fabricated using standard PCB techniques. Each tile measuring \(8.6\times8.6\) cm² (approximately \(8\lambda_0\times8\lambda_0\) at 27.7 GHz) and containing 256 PIN diodes. 

Lastly, the states of the PIN diodes are managed through a dedicated control board integrated onto the rear of the RIS. A four-layer FR4-based control board interfaces with the tiles using 2.5D integration, providing the desired clock and data signals via a microcontroller unit (MCU). Vertical stacking enhances scalability and compactness of the system. Further details on the control board architecture are thoroughly presented in \cite{shekhawat_2025}. While commercial PIN diodes were used for initial proof-of-concept prototype, future designs could adopt transistor based switches for improved power efficiency \cite{venkatesh_2020}.

\subsection{RIS Beamforming Characterization}

To effectively characterize the fabricated four-tile RIS prototype, we developed a fully automated and stable near-field measurement system. The near-field measurement configuration is illustrated in Fig.~\ref{fig:mmWRIS_beam_characterization}a. The RIS is illuminated using a quasi-optical setup, with a transmitting pyramidal horn antenna (gain of 15 dBi) and a receiving waveguide probe antenna (gain of 6 dBi). To avoid obstruction of the transmitted signal, the planar positioner is oriented at a radial angle of $45^\circ$, constraining the measurable angular range to approximately $[20^\circ, 60^\circ]$. The receiving antenna collects data over a planar near-field, following a rectangular grid in the $x$-$y$ plane. The collected near-field data are then processed using a near-field to far-field (NF2FF) transformation to obtain the far-field radiation patterns. The bistatic RCS of the RIS is characterized across a frequency bandwidth from 26 GHz to 28 GHz and a peak RCS performance is observed at approximately 27.2 GHz.

Measured results, depicted in Fig.~\ref{fig:mmWRIS_beam_characterization}b, confirm the effectiveness of the RIS in beam steering, with only the primary reflected beam evident across the entire angular range $[20^\circ, 60^\circ]$ in both azimuth and elevation scanning. In contrast, quantization lobes within the negative angular range $[-20^\circ, -60^\circ]$ were substantially suppressed by more than 10 dB, highlighting the efficacy of the implemented random phase delays. Additionally, the measured half-power beamwidth (HPBW) ranged between $4^\circ$ and $6.5^\circ$, corresponding to reflection angles from $20^\circ$ to $60^\circ$, respectively.


\begin{figure*}[ht]
    \centering
    \includegraphics[width=\textwidth]{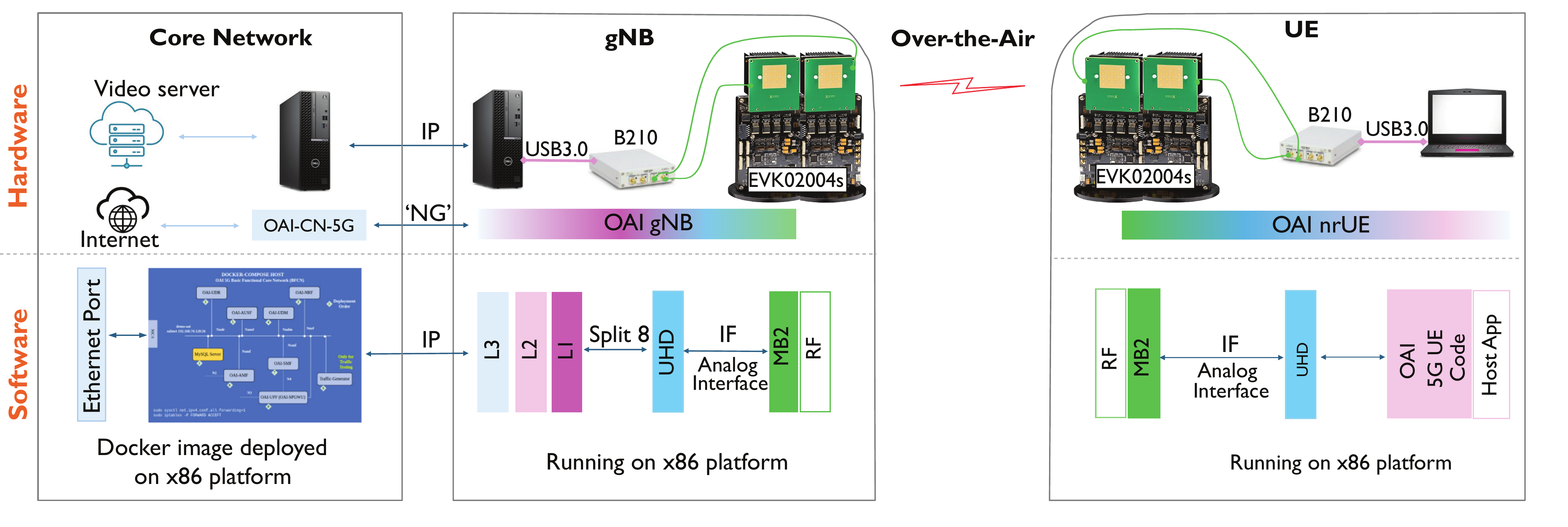}
    \caption{mmWave SDR-based RAN testbed: Detailed depiction of the RAN architecture, including software protocol stacks and their corresponding hardware components.}
    \label{fig:mmWORAN}
\end{figure*}

\section{RIS-Assisted mmWave O-RAN System} \label{sec:proposed_soln}

Integrating RISs into 5G systems introduces several challenges, particularly in scenarios involving user mobility. In this section, we present the proposed framework for integrating RISs into O-RAN systems by leveraging the open and standardized E2 interfaces defined by the O-RAN architecture. Section~\ref{subsec:motivation} outlines the motivation for the proposed platform, highlighting its potential for performance enhancement and operational flexibility. Section~\ref{subsec:problem-formulation} formulates the problem by demonstrating how signal power information made available through the O-RAN interface can be utilized to optimize RIS beam configurations and UE beamforming vectors. Section~\ref{subsec:impl} describes the implementation of the mmWave O-RAN system. Finally, Section~\ref{sec:operation} discusses the operation of the RIS-assisted mmWave O-RAN system, demonstrating its feasibility and practical deployability.

\subsection{Motivation for the Proposed Solution} \label{subsec:motivation}
RISs are recognized as one of the most promising advancements for beyond 5G and 6G wireless networks\cite{Ayub_2021}. Evaluating their performance in practical systems, however, is a critical step towards the standardization and widespread adoption. In this context, one of the main questions is how RIS operation, including beam management, could be carried out within 3GPP systems. Integrating RISs  into existing cellular networks necessitates extensions of current 3GPP beam management standards, as RIS beams must be dynamically adjusted and in coordination with cellular nodes to maintain reliable communication links in mmWave (FR2)  frequency bands.

Given these challenges, there is a compelling need to design and implement RIS-assisted Radio Access Network (RAN) systems that support RIS system integration and facilitate the development of advanced beam management strategies. In this work, we leverage the standardized interfaces provided by O-RAN to seamlessly integrate RISs into a 3GPP-compliant RAN. We then use this testbed to develop and evaluate mobility-aware beam management and coverage enhancement experimentation.

\subsection{Problem Formulation} \label{subsec:problem-formulation}
In this work, we consider the case where the direct link between the gNB and the UE is completely blocked, and the received signal power from multipath links is negligible. To maximize the signal-to-noise ratio (SNR) at the UE, we assume a fixed transmit beamforming vector at the gNB, as both the gNB and RIS are stationary, while a joint exhaustive beam search is performed at the UE and the RIS. The receive combining vectors, \( \mathbf{w}_\text{UE}^{m} \in \mathbb{C}^{1 \times M_\text{UE}} \), are selected from a predefined codebook \(\boldsymbol{ \mathcal{W}} = \{\mathbf{w}_\text{UE}^{m}\}_{m=1}^{M_\text{UE}} \). The RIS interaction vector \( \boldsymbol{\psi}_{\text{RIS}} \) is chosen from a predefined codebook \( \boldsymbol{ \mathcal{P} }\), assuming that the RIS radiating elements act as RF phase shifters. At each UE location, the optimal beamforming and interaction vectors, \( \{\mathbf{w_\text{UE}}^\star, \boldsymbol{\psi}_{\text{RIS}}^\star\} \), are determined as \eqref{eq:snr_equation}\:
\begin{equation}
\{\mathbf{w_\text{UE}}^\star, \boldsymbol{\psi}_{\text{RIS}}^\star\} = 
\argmax_{\mathbf{w_\text{UE}} \in \mathcal{W},~ \boldsymbol{\psi_{\text{RIS}}} \in \mathcal{P}} 
\left| \mathbf{w_\text{UE}}^H \bar{\mathbf{H}} \mathbf{f}_{\text{gNB}}  \right|,
\label{eq:snr_equation}
\end{equation}
where \( \bar{\mathbf{H}} = \mathbf{H}_{\text{RIS,UE}} \mathbf{\Psi}_{\text{RIS}} \mathbf{H}_{\text{gNB,RIS}} \) is the cascaded channel matrix. The sets \( \boldsymbol{ \mathcal{W}} \) and \( \boldsymbol{ \mathcal{P}} \) represent the combining vectors used by the UE and the codewords of RIS elements, respectively.                             

We investigated the use of received signal power at the UE to determine the optimal parameters,  \( \{\mathbf{w_\text{UE}}^\star, \boldsymbol{\psi}_{\text{RIS}}^\star\} \) required to ensure a reliable mmWave communication link in a private 5G network. Additionally, the standard open interfaces provided by the O-RAN architecture were leveraged to monitor the UE signal strength and facilitate adaptive beamforming and control of RIS interaction vectors in dynamic scenarios.

\subsection{Testbed Implementation} \label{subsec:impl}

We implemented an O-RAN-based 5G system operating in the mmWave (FR2) band, as illustrated in Fig.~\ref{fig:mmWORAN}. The testbed comprises three primary subsystems: the core network (CN), the radio access network (RAN), and the near-real-time RAN Intelligent Controller (near-RT RIC). All components are co-located on a single computing platform, while the user equipment (UE) operates independently on a separate machine. The CN, RAN, and UE are realized using the OpenAirInterface (OAI) software stack~\cite{oai}, an open-source SDR framework tailored for prototyping LTE/5G and O-RAN-compliant networks. The hardware specifications of the deployed system are summarized in Table~\ref{fig:components}.

To support mmWave operation, baseband/intermediate-frequency (IF) signals from USRP SDRs are interfaced with EVK02004 mmWave phased array modules, which perform frequency up/down-conversion and beamforming within FR2 bands. This configuration required modifications to the OAI codebase to enable accurate gain control and frequency synchronization between the USRPs and the EVK02004 phased array modules. The testbed is designed to be scalable, and it can support wider bandwidths and larger subcarrier spacing via hardware upgrades (e.g., replacing B210s with X310/X410 SDRs). For improved modularity and portability in UE mobility use cases, the baseband stack of the mmWave UE—originally composed of the OAI-based \texttt{nr-uesoftmodem} and USRP B210, can be substituted with a commercial off-the-shelf (COTS) UE, such as the Quectel RM500-GL module.

The near-RT RAN control is enabled through the integration of FlexRIC framework~\cite{flexric} with the OAI-RAN. We modified the standard E2SM-KPM implementation in FlexRIC to extract additional key performance indicators (KPIs), including the reference signal received power (RSRP) and the UE’s unique identifier. We developed and deployed an xApp program on the FlexRIC platform. The xApp periodically monitors the RSRP values reported by the UE via the standardized E2 interface and dynamically manages the RIS and mmWave phased array beamforming configurations based on observed RSRP trends. To configure the RIS beam state, the xApp communicates with a TCP server hosted on a remote machine or cloud instance (e.g., AWS EC2). This server transmits RIS beam index commands to a TCP client program running on the RIS controller (MCU), which selects and activates the codewords corresponding to the received beam index. In parallel, a separate TCP client program is implemented on the gNB and UE compute units to monitor and update the mmWave phased array beams based on control commands received from the remote/cloud-based TCP server.

The implemented testbed platform enables seamless integration of the RIS into a live, O-RAN-compliant 5G system. It supports low-latency RIS control, making it well-suited for fast beam adaptation under UE mobility. The platform further facilitates experimental validation of RIS-assisted communication algorithms in realistic deployment environments. Additional implementation and operational details are provided in Section~\ref{sec:operation}.

\begin{table}[t]
    \centering
    \caption{Hardware components and specification}
    \label{fig:components}
    \resizebox{\columnwidth}{!}{%
    \begin{tabular}{|>{\raggedright\arraybackslash}m{2.85cm}|>{\raggedright\arraybackslash}m{5cm}|}
        \hline
        \textbf{Components} & \textbf{Specifications} \\ \hline

        \multicolumn{2}{|l|}{\textbf{gNB}} \\ \hline
        Desktop computing unit & Dell Precision 3680 Tower; Intel Core @ 3.4\,GHz; 32\,GB RAM; 20 cores (28 threads) \\ \hline
        Baseband Radio & NI USRP B210, interfaced over USB 3.0 \\ \hline
        mmWave RF-frontend & Two EVK02004 units, each with a $4\times4$ UPA (16 elements) and 32 beamforming channels; supporting dual polarization \\ \hline

        \multicolumn{2}{|l|}{\textbf{UE}} \\ \hline
        Laptop computing unit & Alienware m15 R7; Intel Core 12th Gen i9 @ 5.0\,GHz; 64\,GB RAM; 20 cores (2 threads/core) \\ \hline
        Baseband Radio & NI USRP B210 (USB 3.0) or Quectel RM500-GL with two-way RF power splitter/combiner \\ \hline
        mmWave RF-frontend & Same RF modules as gNB \\ \hline

        \multicolumn{2}{|l|}{\textbf{mmWave RIS}} \\ \hline
        Power \& Computing unit & Jetson Xavier NX, Arduino DUE, and 8\,W power supply \\ \hline
        Electromagnetic properties & Linear polarization; center frequency 27.2\,GHz; HPBW $4^\circ$ – $6.5^\circ$, scanning angles  +/- $60^\circ$ \\ \hline
    \end{tabular}
    }%
\end{table}

\begin{figure*}[h!]
    \centering
    \includegraphics[width=.65\textwidth]{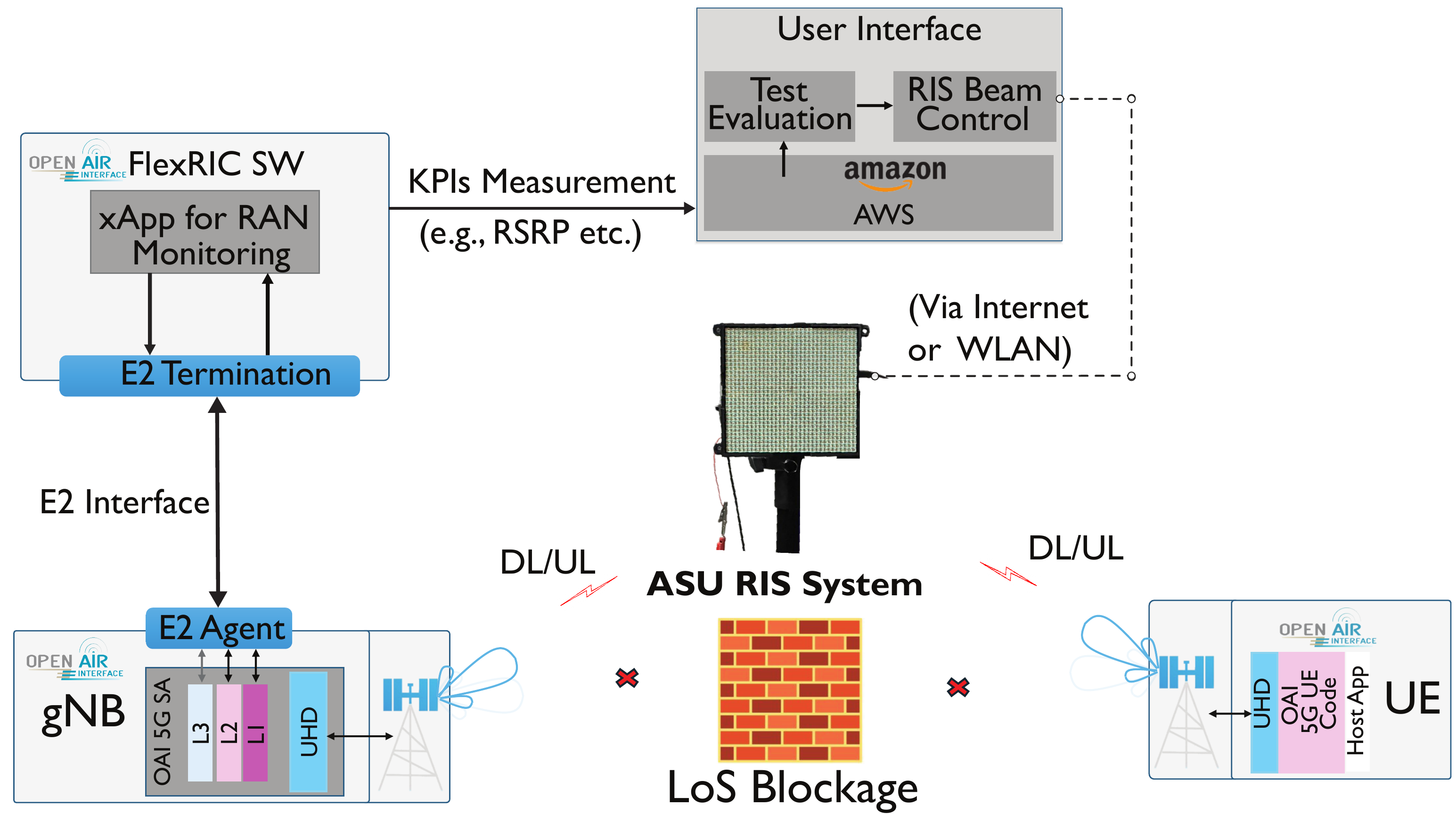}
    \caption{Illustration of the proposed O-RAN and RIS integration framework. The figure shows how an xApp, deployed on FlexRIC, monitors key performance indicators (KPIs), which are then used to dynamically configure the RIS. The RIS reflects and steers signals between the gNB and the UE to enhance link reliability in FR2 frequency bands.}
    \label{fig:platform}
    \vspace{-2mm}
\end{figure*}

\subsection{End-to-End Testbed Operation}\label{sec:operation}
This section outlines the operational workflow of the implemented testbed, which comprises multiple subsystems distributed across dedicated computing units. The process begins with the initialization of the mmWave RF front-ends, followed by the launch of the RIS controller script, which interfaces with a TCP server hosted on a designated computing unit: a remote machine or a cloud instance. Next, the core 5G software components (\texttt{OAI-gNB}, \texttt{OAI-nrUE}, the near-RT RIC, and the xApp) are executed to establish UE connectivity, enable real-time RIS control, and facilitate beam management within the integrated O-RAN system.

\subsubsection{mmWave RF Front-End Modules}

The first step in the testbed operation involves configuring the mmWave RF front-end modules to establish directional links between the gNB, RIS, and UE. We leverage a software API (\texttt{mb2}) provided by Sivers Semiconductors to initialize the EVK02004 mmWave phased array modules in either transmit (TX) or receive (RX) mode, depending on their roles in the time-division duplexing (TDD) configuration described in Table~\ref{tab:oai_features}. In addition, we developed custom control functions on top of the \texttt{mb2} API to enable real-time beam configuration and gain adjustment of the phased array modules.

We configure the beamforming coefficients of the phased array elements using beamforming vectors selected from a predefined codebook. At the gNB, the beam is typically set to the boresight direction ($\theta = \phi = 0^\circ$), while at the UE, the beam is directed toward the RIS, with adjustments based on the location of the UE. For the indoor and outdoor scenarios shown in Fig.~\ref{fig:ris-indoors-outdoors}, the gNB is positioned at the RIS boresight, and the UE array is tilted to align its beam with the RIS center, maximizing the received signal strength at the initial UE location. To optimize link performance under real-world conditions, we also adjust the power amplifier (PA) and low-noise amplifier (LNA) settings in the TX and RX chains of the RF front-end modules. In addition, USRP gain configurations are tuned accordingly. These gain adjustments help ensure robust signal transmission and reception across varying environmental conditions and hardware setups.

\subsubsection{Interfacing Server with RIS Controller}

Following the initialization of the mmWave RF front-ends at both the gNB and UE, a TCP server program is executed on a designated computing unit: a remote machine or a cloud-based instance. This server interfaces with the xApp to periodically receive key performance indicators (KPIs) and send control commands to the RIS controller based on real-time system conditions. The server hosts several key control logics, including: (i) initiating RIS beam sweeping when the UE is disconnected from the gNB due to poor coverage; (ii) coordinating RIS beam switching in response to UE mobility after a stable link has been established; (iii) refining the RIS beam by probing neighboring codewords to maximize the received signal power at the UE; and (iv) triggering UE mobility management procedures. The RIS controller, implemented on an MCU, runs lightweight firmware that continuously listens for control commands over a serial link(e.g., UART/USB). These commands are interpreted as updated RIS beam configurations received from either the TCP server or a user interface platform, as discussed in subsequent sections.

\subsubsection{OAI-RAN and Near-RT RIC Programs}

The next step in the system operation involves launching the near-RT RIC, \texttt{OAI-gNB}, \texttt{OAI-nrUE}, and the custom xApp programs. In our implementation, the \texttt{OAI-gNB} and \texttt{OAI-nrUE} are compiled and executed on separate computing units, as detailed in Table~\ref{fig:components}. The near-RT RIC is co-located with the \texttt{OAI-gNB} on the same machine, although it can alternatively be hosted on a separate server if required by the deployment.

The near RT RIC is initialized first to activate the service ports necessary for communication with both the RAN and the xApp. Subsequently, the \texttt{nr\_softmodem} and \texttt{nr\_uesoftmodem} binaries, responsible for executing the \texttt{OAI-gNB} and \texttt{OAI-nrUE} stacks, respectively, are launched with appropriate configuration files and runtime parameters. This sequence establishes the RAN–UE connection, enables real-time KPI reporting from the UE to the xApp via the E2 interface, and facilitates RIS control through the TCP server. The key features and operational parameters of the OAI-based mmWave O-RAN testbed are summarized in Table~\ref{tab:oai_features}.

\subsubsection{User Interface}

As the final component in the operational pipeline, we developed a graphical user interface (GUI) that interfaces with both the TCP server and the xApp. This GUI provides real-time visualization of key performance indicators (KPIs), including the reference signal received power (RSRP) and the currently active RIS beam index. Beyond monitoring, the GUI enables beam switching control for the gNB, UE, and RIS modules, based on real-time RSRP feedback. This capability complements the autonomous beam control logic handled by the xApp and TCP server, while also supporting manual override and testing scenarios.

Furthermore, this GUI functionality can be modularized and deployed as a dedicated xApp microservice on the near-RT RIC, thereby enabling scalable and low-latency control over mmWave transceiver elements. This design enhances both observability and flexibility in managing dynamic wireless environments.

\begin{table}[h!]
    \centering
     \caption{Main features of the OAI deployment.}
    \resizebox{\columnwidth}{!}{%
    
    \begin{tabular}{l@{\hspace{1.5cm}}l@{\hspace{1.0cm}}}
	 \thickhline
       Feature & Description \\ 
	 \thickhline
        3GPP Release & 15 \\ 
        Carrier Frequency & 27 GHz ($@$ Band n257) \\ 
        Band Type & TDD \\
        Intermediate Frequency & 5.0 GHz (or $@$ Band 78) \\
        Bandwidth & 40 MHz \\ 
        Subcarrier Spacing & 30 kHz \\ 
        TDD Period  & 5 ms \\
        TDD Slot Configuration & DDDDDDDFUU \\
        SSB Periodicity &  20 ms \\   \thickhline
    \end{tabular}
    }

    \label{tab:oai_features}
\end{table}

\begin{figure*}[h!]
	\centering
	\subfigure[Indoor Scenario: Experimental setup and geometry]{ \includegraphics[width=.47\textwidth]{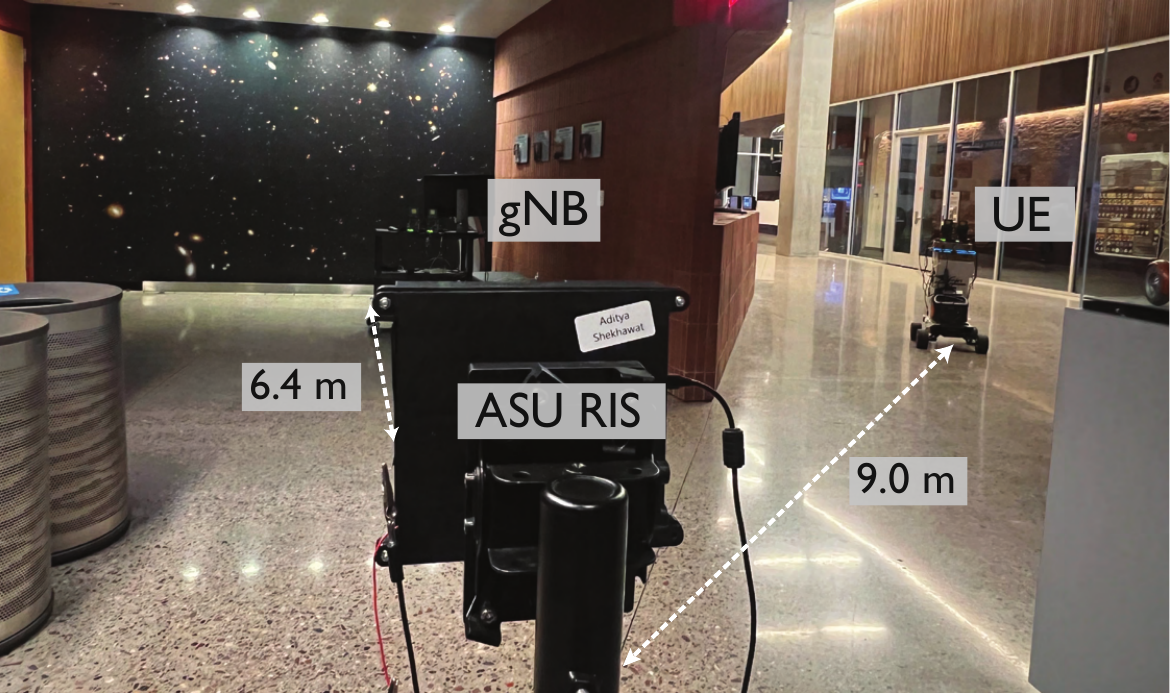} \label{indoor-setups} }\hspace{2pt}
    \subfigure[Outdoor Scenario: Experimental setup and geometry]{ \includegraphics[width=.42\textwidth]{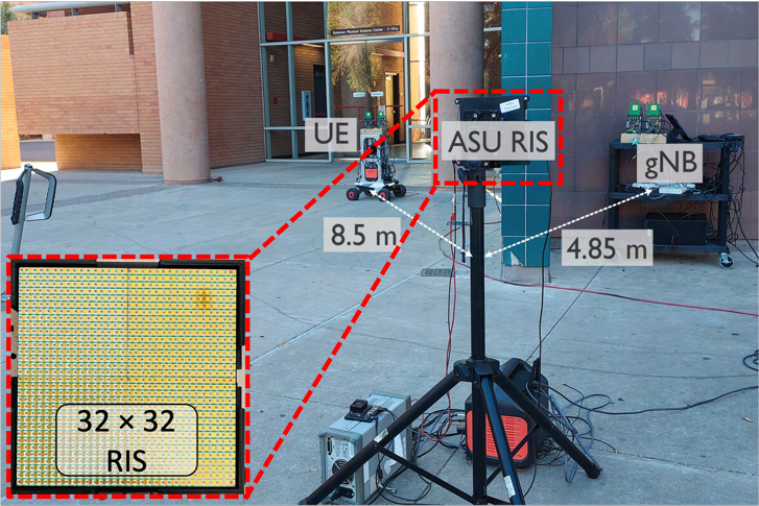} \label{outdoor-setups} }
\caption{ Field trial sites used for coverage data collection and evaluation of UE mobility management algorithms in both indoor and outdoor environments.}
	\label{fig:ris-indoors-outdoors}
\end{figure*}

\begin{figure*}[h!]
	\centering
	\subfigure[Indoor Scenario: Floor-view and UE Trajectory]{ \includegraphics[width=.42\textwidth]{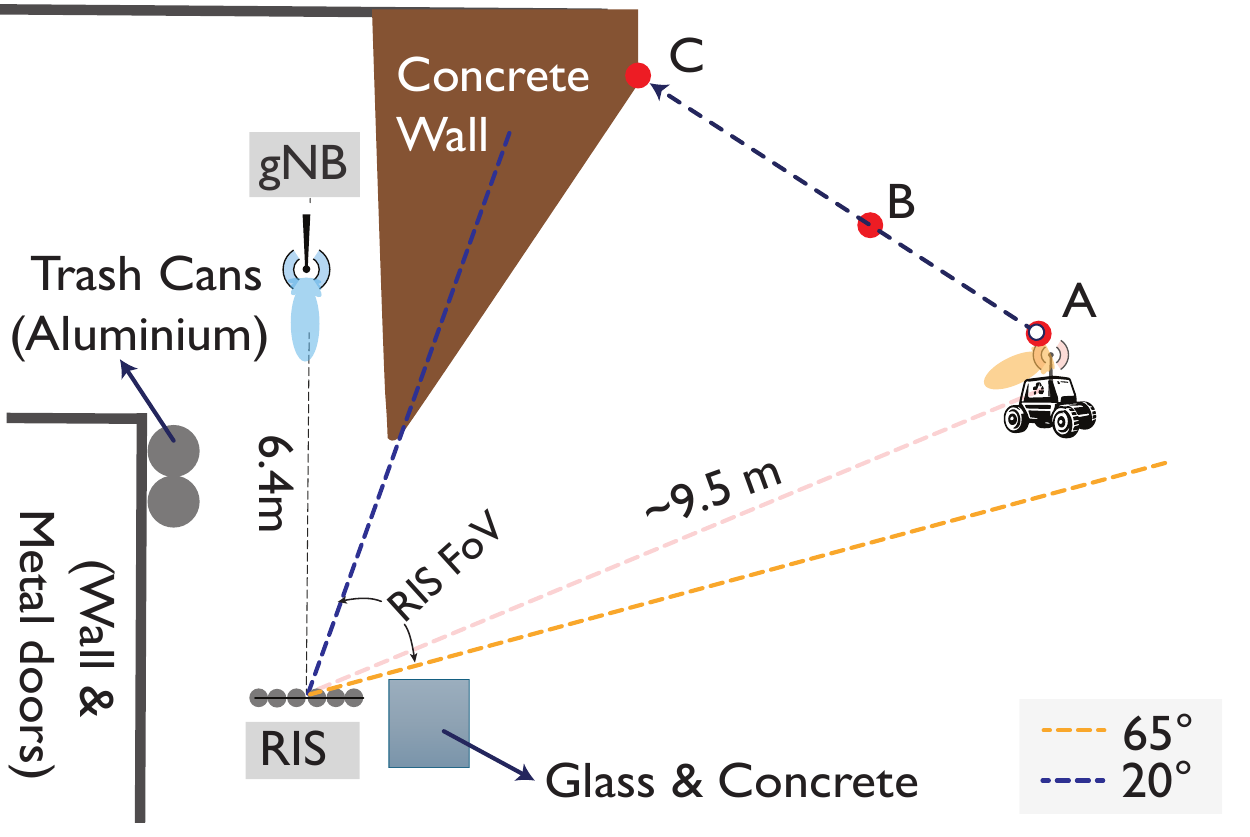} \label{indoor-traj} }\hspace{2pt}
    \subfigure[Outdoor Scenario: Satellite-view and UE Trajectory]{ \includegraphics[width=.42\textwidth]{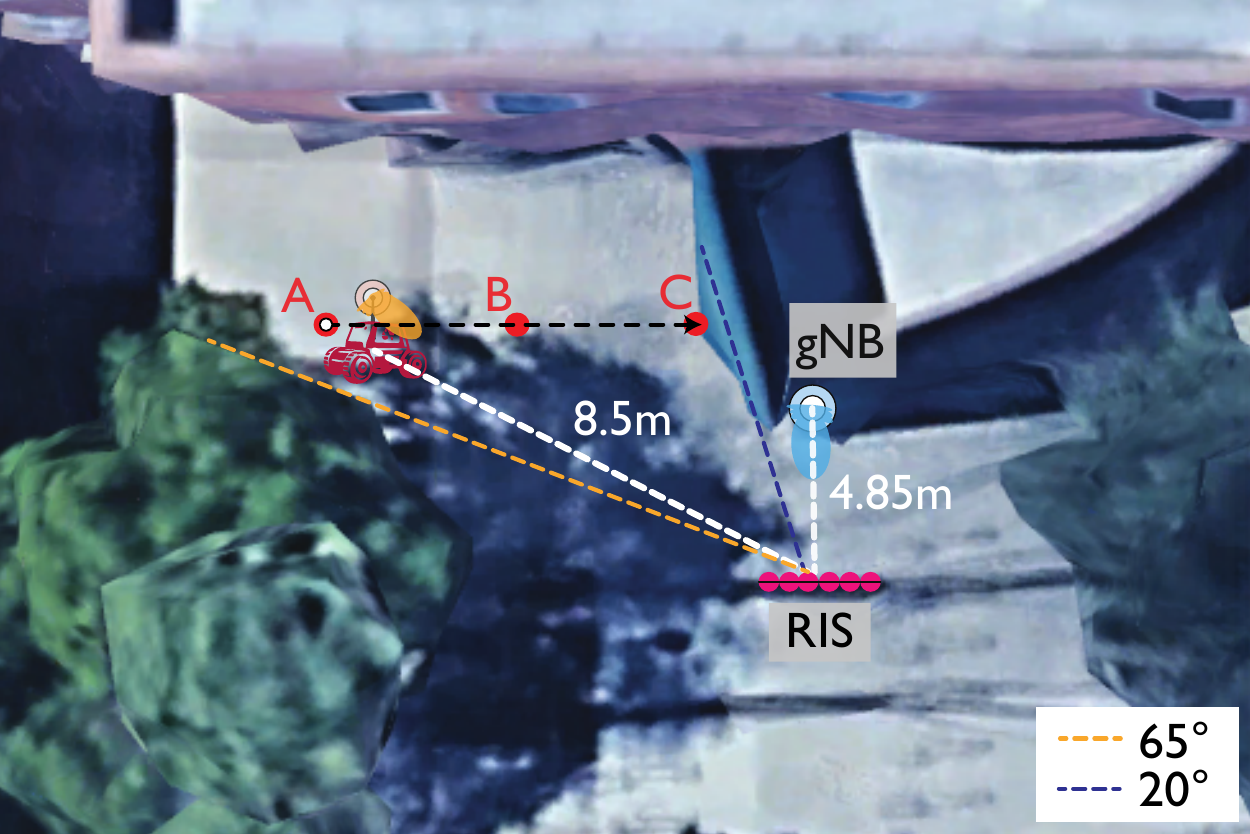} \label{outdoor-traj} }

\caption{Mobility trajectories used to evaluate the performance of the proposed mobility management algorithms. The UE module, installed on an automated robot, moves from point A to point C. In some trials, the robot passes through point B; in others, it moves from A to B, returns to A, and then proceeds to C.}
	\label{fig:trajectory}
\end{figure*}

\begin{figure*}[ht]
    \centering
    
   \subfigure[ RSRP without RIS]{ \includegraphics[width=0.3\textwidth]{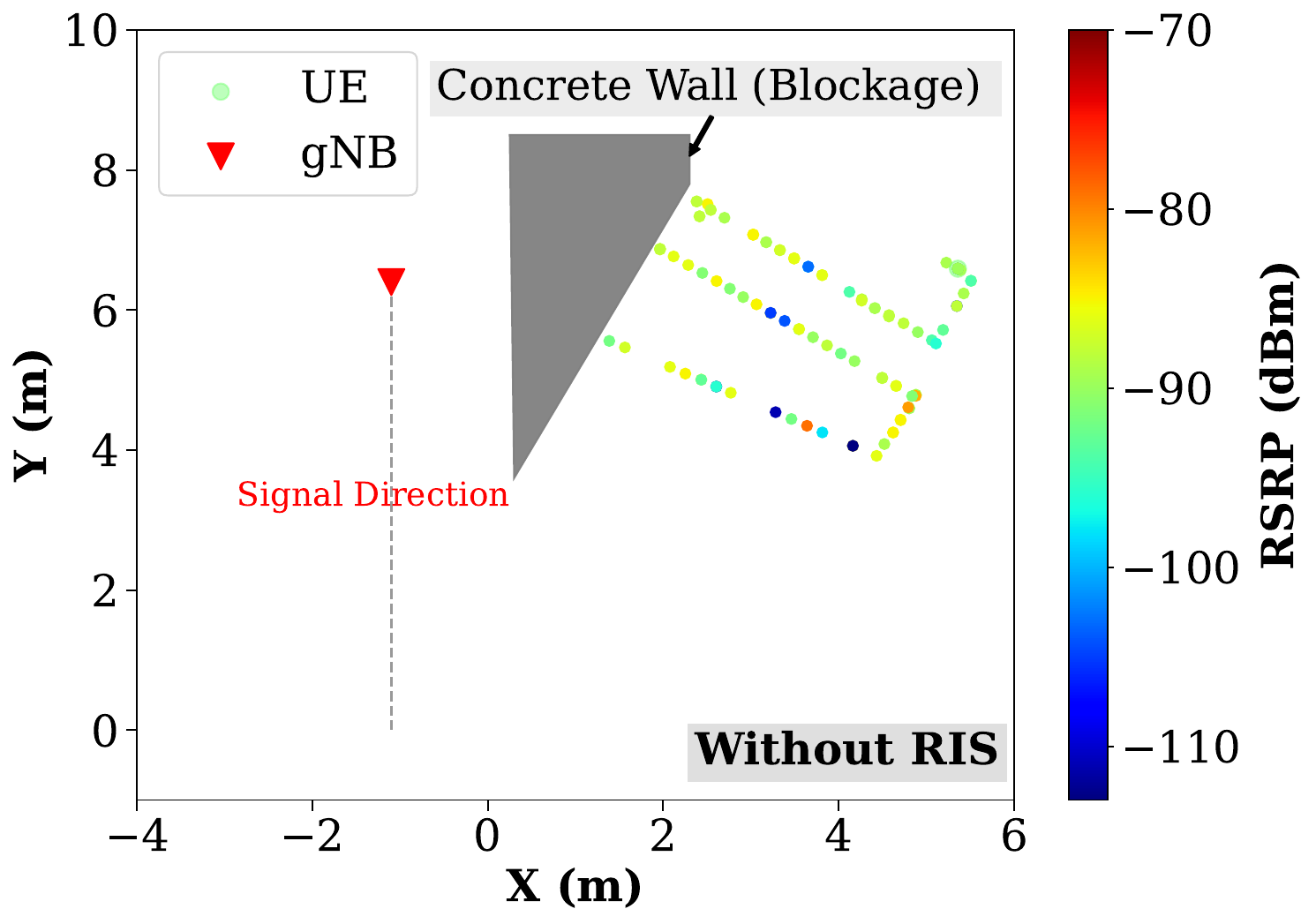} \label{coverage:1a} }
    \subfigure[  RSRP with RIS]{ \includegraphics[width=0.3\textwidth]{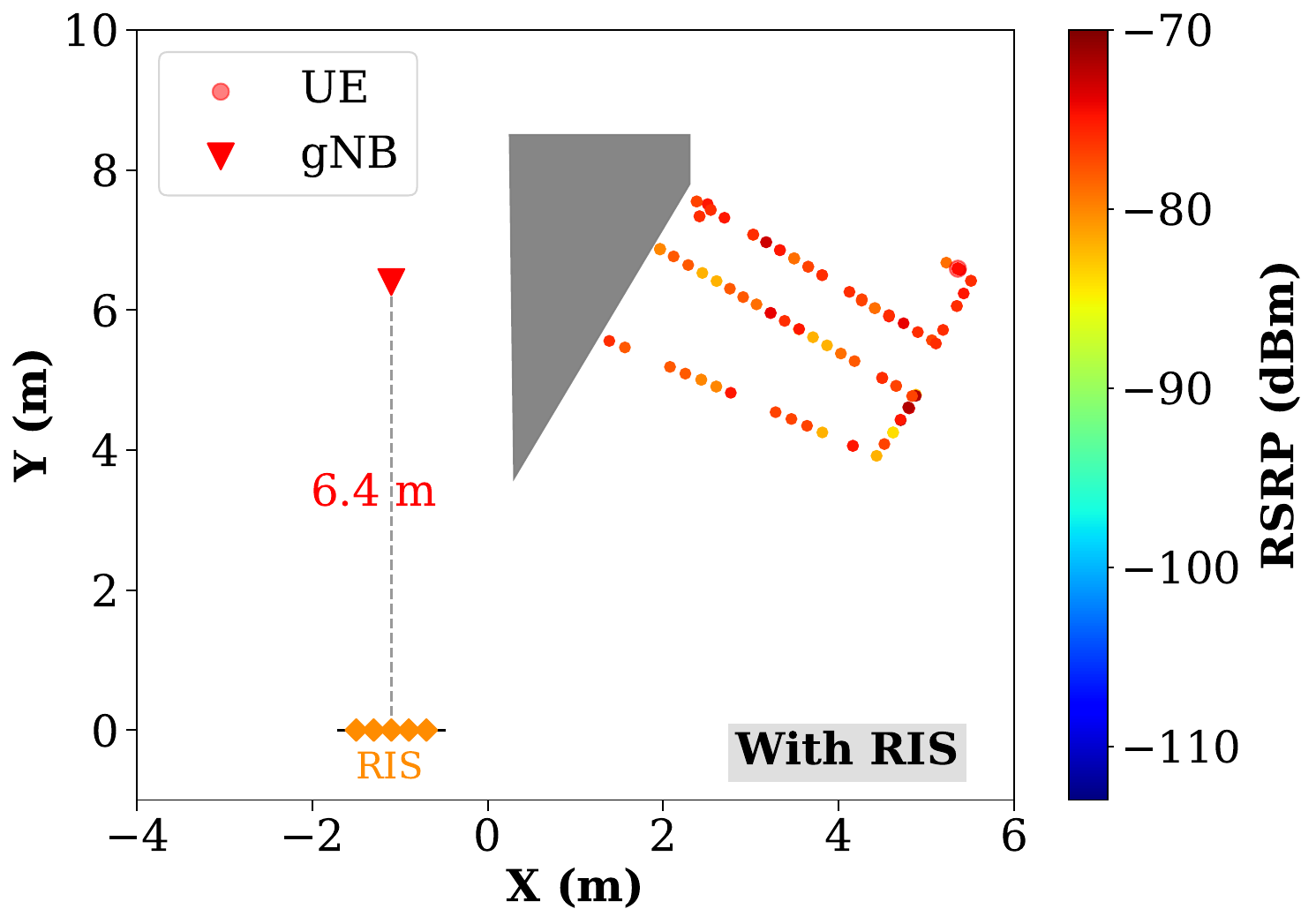} \label{coverage:1b} }
    \subfigure[ RSRP Gain with RIS]{ \includegraphics[width=0.3\textwidth]{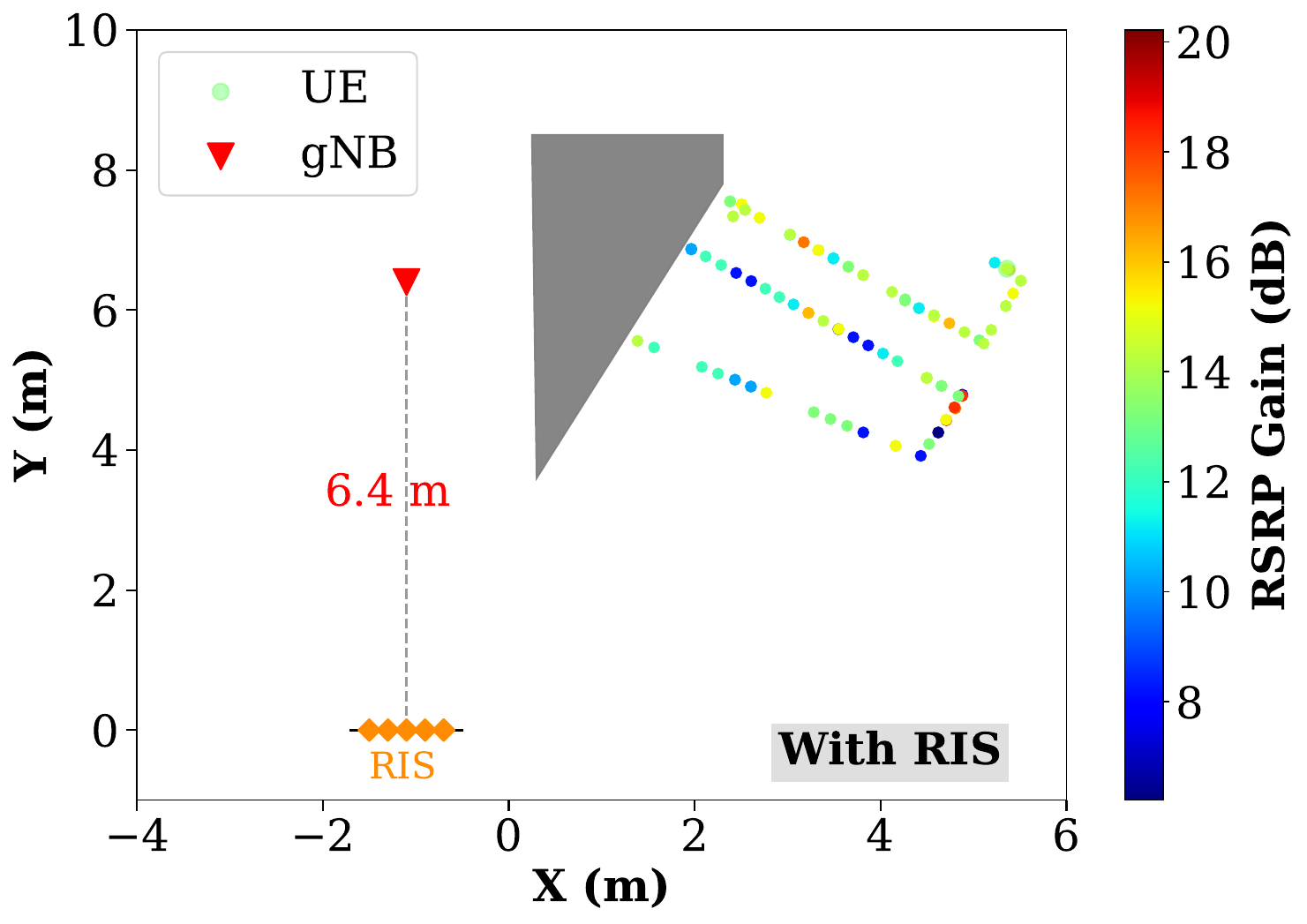} \label{coverage:1c} }

    \subfigure[ Throughput without RIS]{ \includegraphics[width=0.3\textwidth]{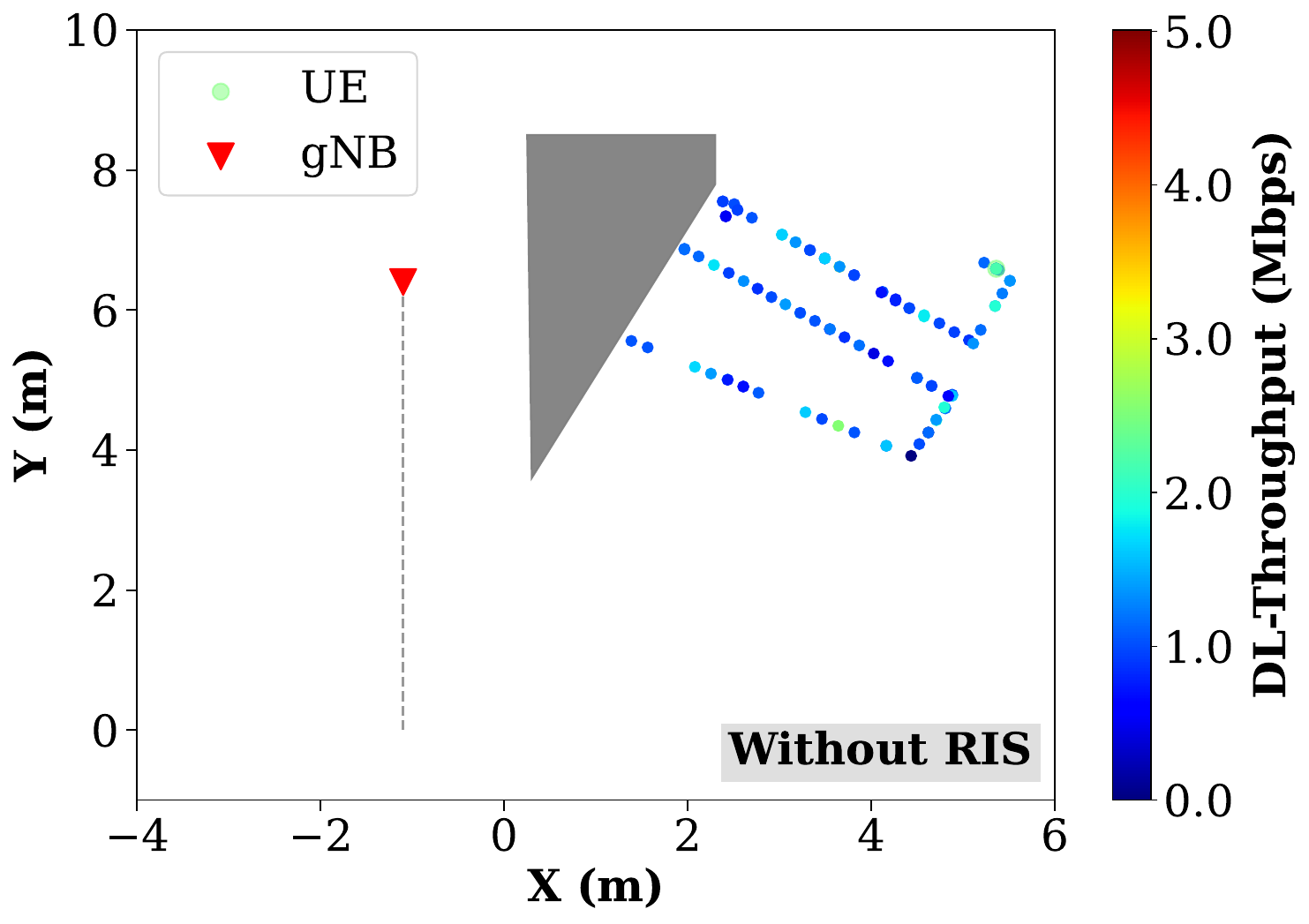}} \label{coverage:2a} 
    \subfigure[ Throughput with RIS]{ \includegraphics[width=0.3\textwidth]{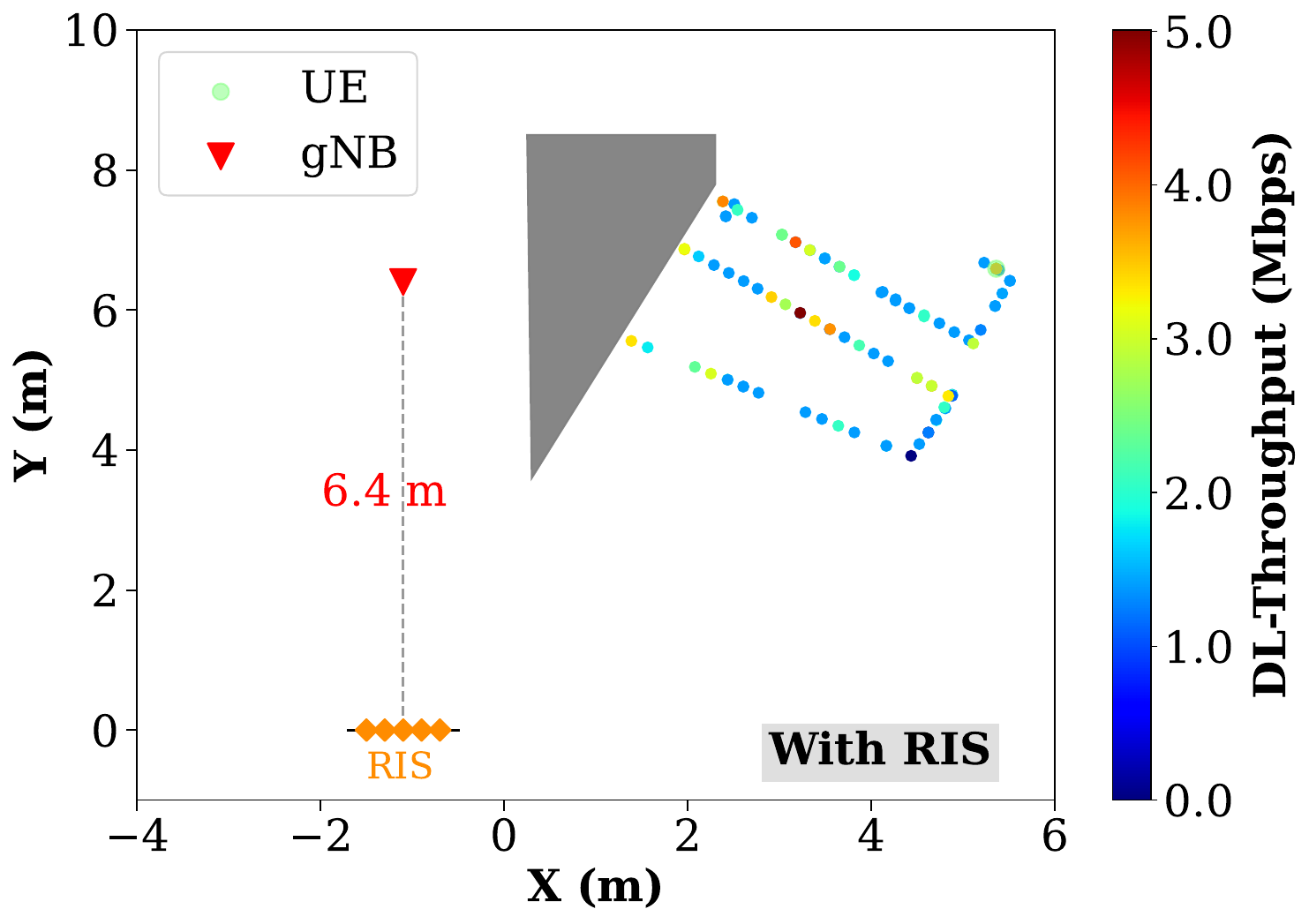} \label{coverage:2b} }
    \subfigure[ Throughput Gain with RIS]{ \includegraphics[width=0.3\textwidth]{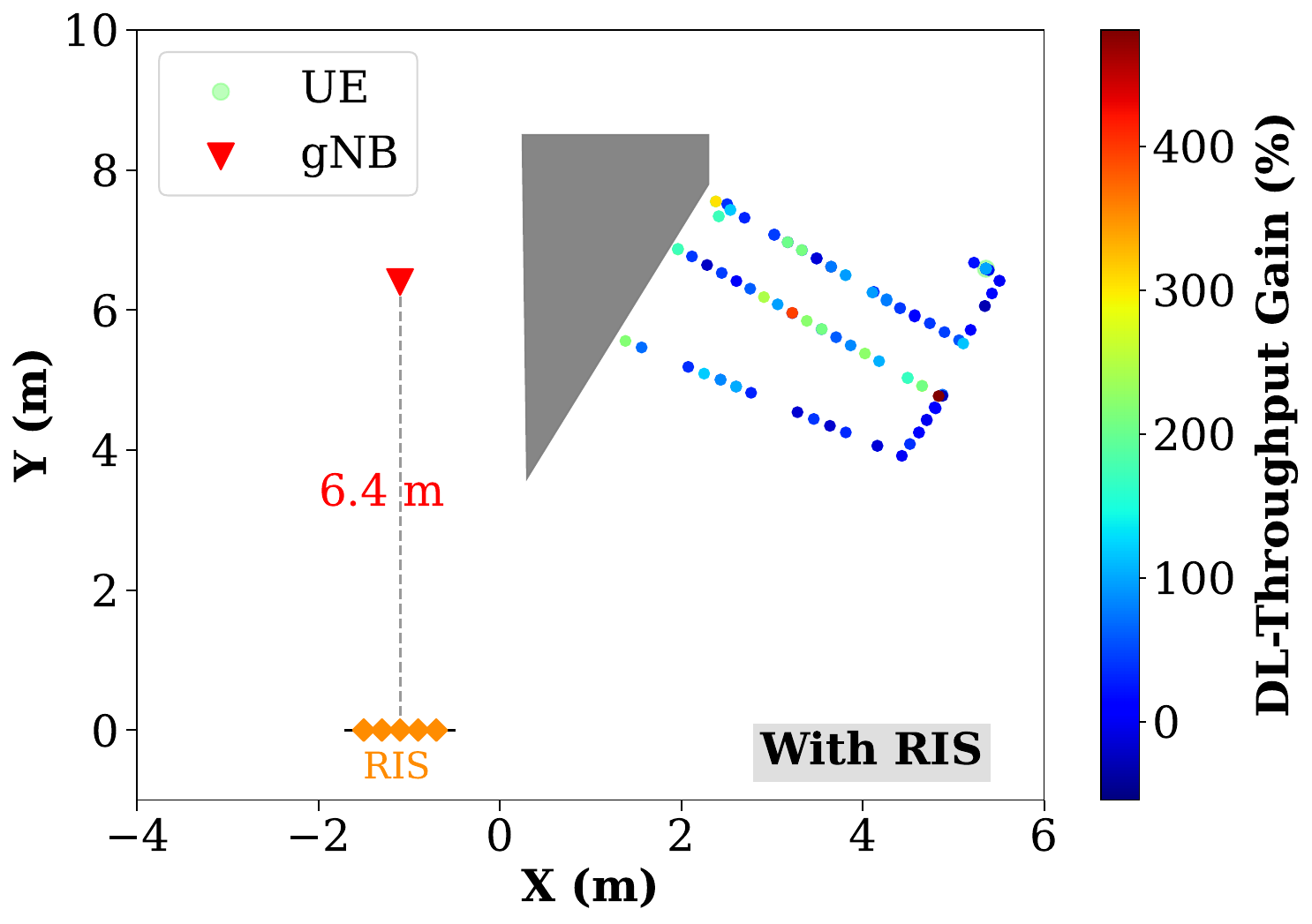} \label{coverage:2c} }

    \subfigure[ Link KPIs]{ \includegraphics[width=0.45\textwidth]{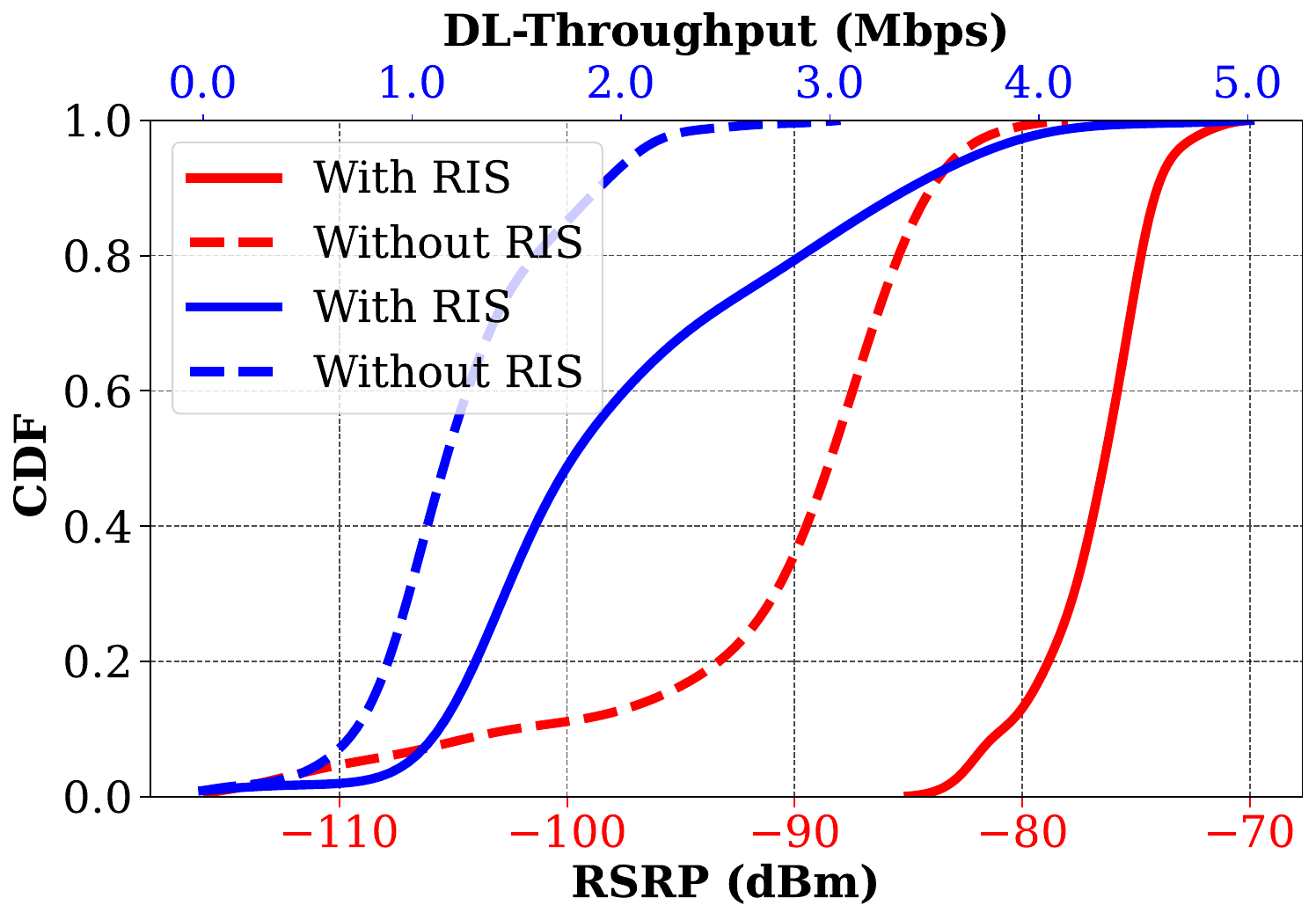}} \label{coverage:3a}     
    \subfigure[ KPI Gains with RIS]{ \includegraphics[width=0.45\textwidth]{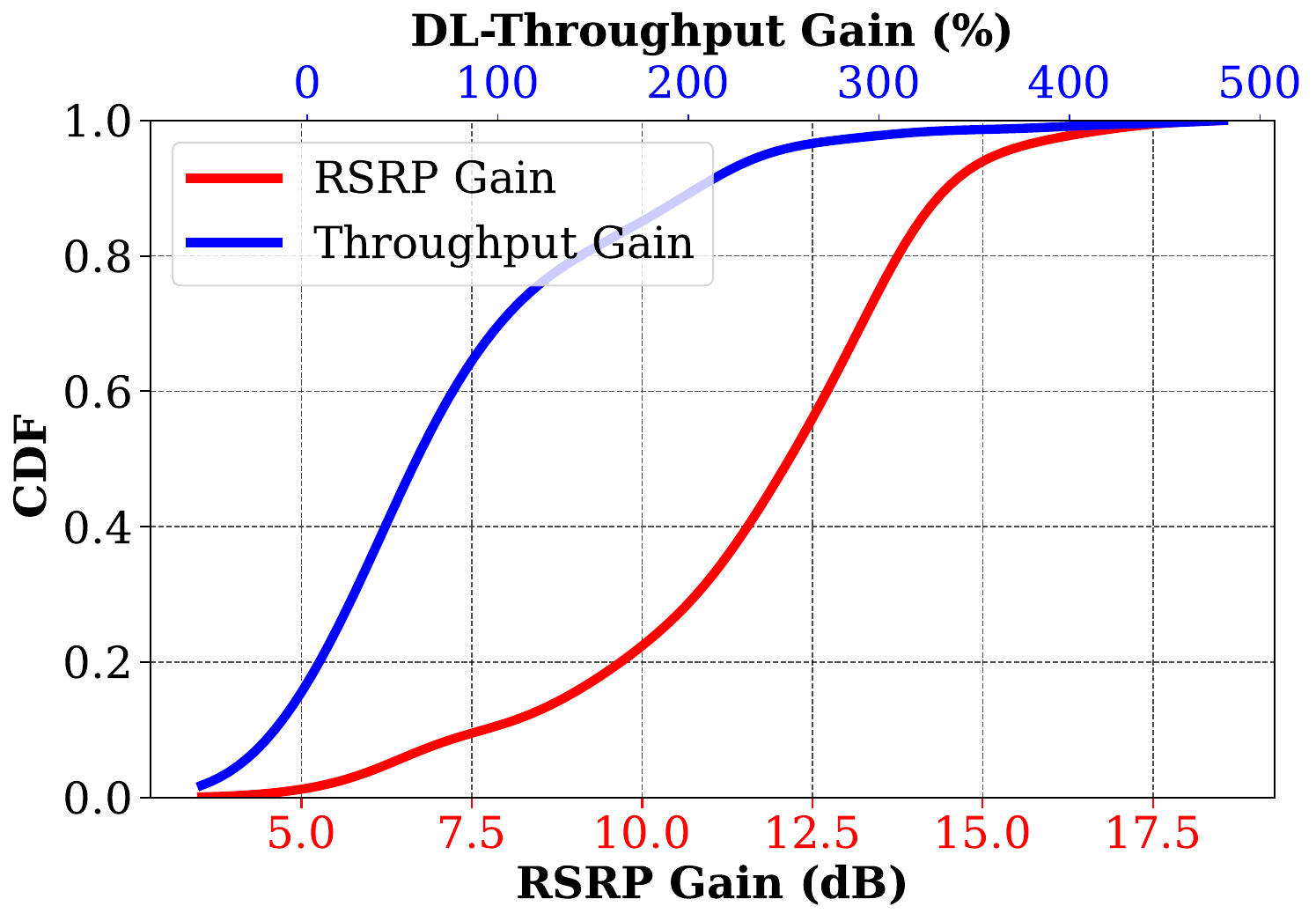} \label{coverage:3b} } 
    
   \caption{Coverage enhancement results for the indoor scenario illustrated in Fig.~\ref{indoor-setups}.}

    \label{fig:coverage_indoors}
    \vspace{-2mm}
\end{figure*}

\begin{figure*}[h]
    \centering
    \subfigure[ RSRP without RIS]{ \includegraphics[width=0.32\textwidth]{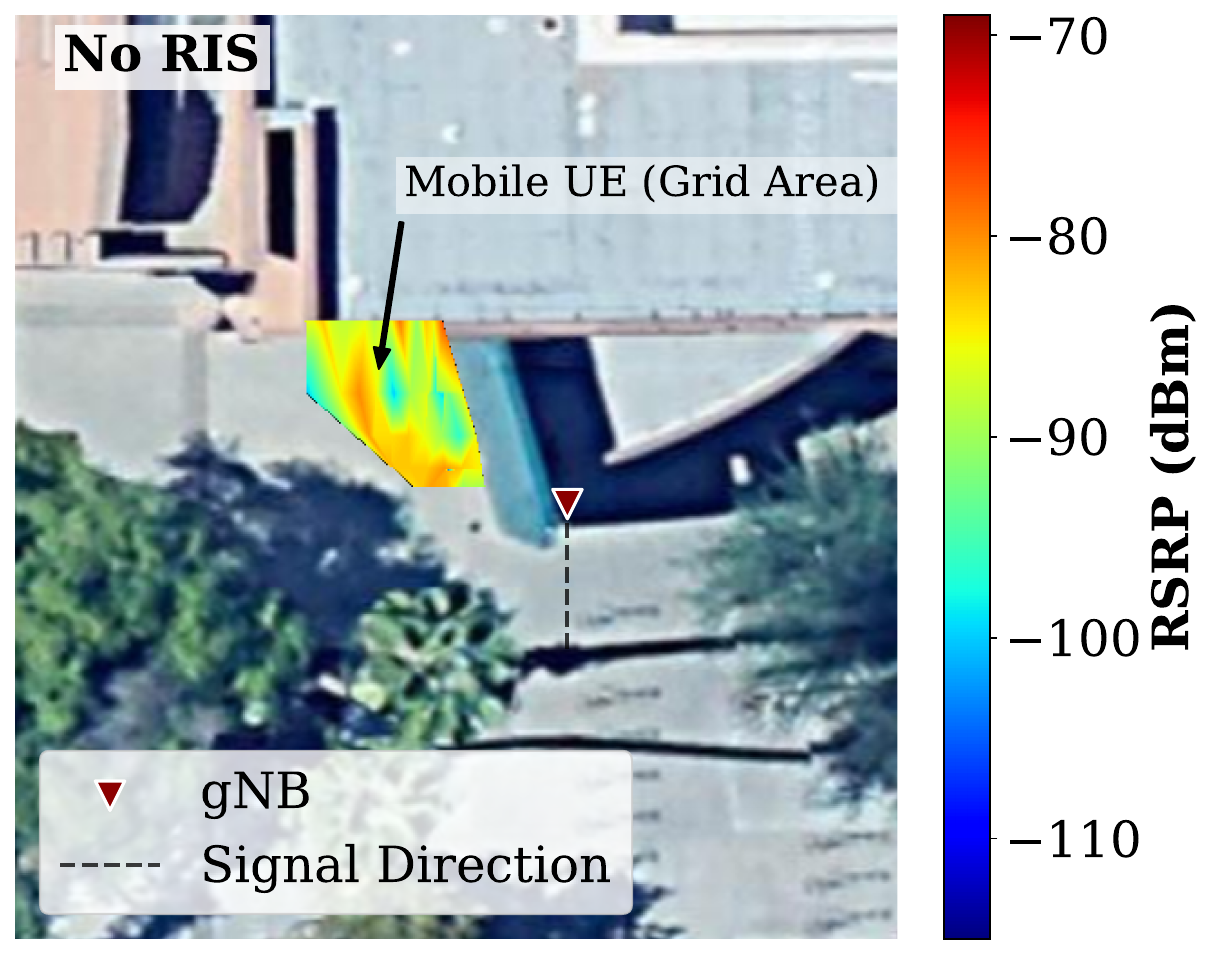} \label{coverage:a} }
    \subfigure[  RSRP with RIS]{ \includegraphics[width=0.31\textwidth]{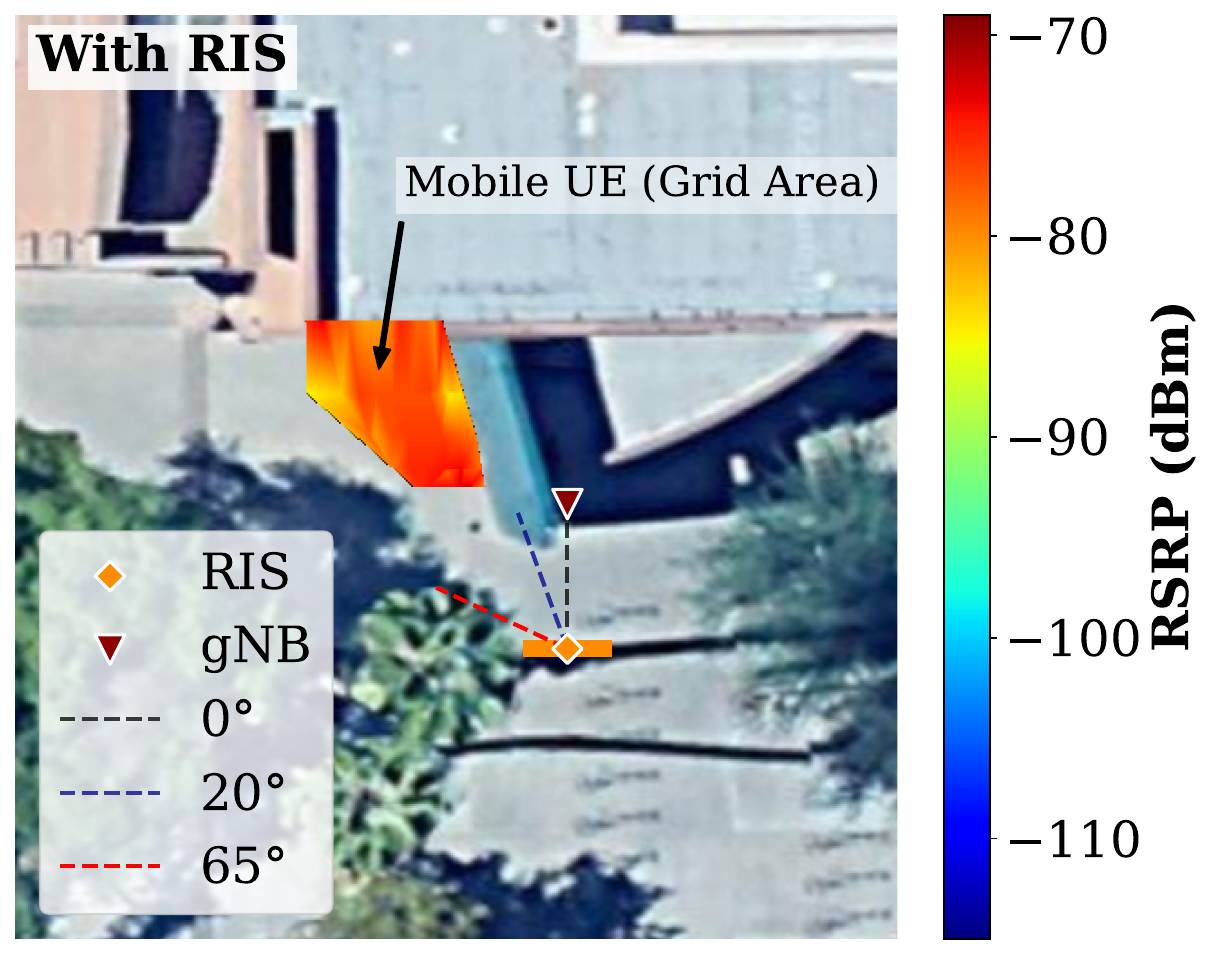} \label{coverage:b} }
    \subfigure[ RSRP Gain with RIS]{ \includegraphics[width=0.295\textwidth]{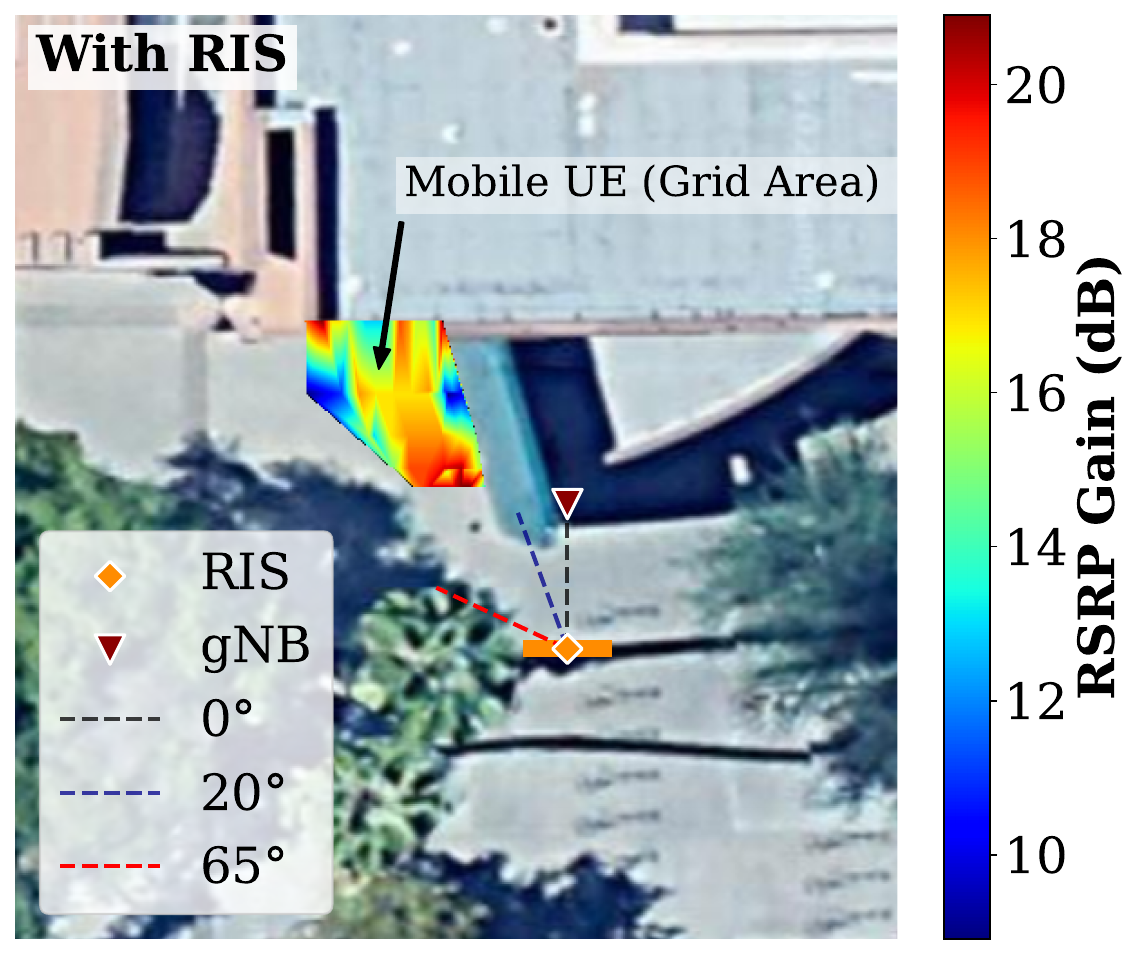} \label{coverage:c} }
   \caption{Coverage enhancement results for the outdoor scenario shown in Fig.~\ref{outdoor-setups}, where the gNB and UE are separated by a concrete structure.}
    \label{fig:coverage_outdoors}
    \vspace{-2mm}
\end{figure*}

\section{Field Trial Scenarios} \label{sec:field}
We conducted a comprehensive set of indoor and outdoor field measurements using the experimental setups illustrated in Fig.~\ref{fig:ris-indoors-outdoors}. In these scenarios, a base station (gNB) equipped with a fixed directional beam was positioned a few meters from the RIS. A mobile mmWave user followed predefined trajectories within the shadowed region, where signals reflected by the RIS dominate due to the obstruction of direct paths by concrete structures.

The system framework, shown in Fig.~\ref{fig:platform}, was configured to capture key performance metrics, including the reference signal received power (RSRP) and downlink(DL) throughput at the UE. These metrics are critical for evaluating signal quality, coverage, and capacity. Fig.~\ref{fig:ris-indoors-outdoors} highlights essential features of the deployment, such as line-of-sight obstructions caused by concrete walls and reflections from nearby surfaces, which can lead to significant multipath signal towards UE. Field measurements were collected at multiple UE positions using different RIS beam configurations. The resulting dataset provides a robust foundation for evaluating system performance across a range of deployment and environmental conditions.

\subsection{Indoor Scenario} \label{sec:indoor}

The indoor trials were conducted on the first floor of ASU's Interdisciplinary Science \& Technology Building IV. This sceario features reflective surfaces such as glass, metal cans and concrete walls, as shown in Fig.~\ref{indoor-setups}. In this setup, the gNB and UE were separated by a concrete wall, resulting in severe signal blockage and high penetration loss. In most shadowed locations, the UE could not detect gNB signals. In some areas, weak reflections enabled limited signal reception, but the corresponding RSRP values remained low and link stability was poor. To mitigate these issues, the mmWave RIS system was deployed to enhance signal coverage and improve signal-to-noise ratio(SNR). By redirecting reflected signals into the shadowed region, the RIS enabled reliable communication between the UE and gNB, significantly stabilizing RSRP in previously inaccessible positions.

As shown in Fig.~\ref{indoor-setups}, the gNB was placed approximately 6.4 meters from the RIS and beamformed toward its broadside. The RF frontend and beamforming gains of the mmWave phased array modules were optimized to ensure strong signal incidence on the RIS surface. For localization, Marvelmind indoor positioning sensors were used to track the 2D coordinates of the mobile UE. This positional data was used to generate coverage visualizations (Fig.~\ref{fig:coverage_indoors}) and support context-aware analysis of RIS and UE beam management.

\subsection{Outdoor Scenario} \label{sec:outdoor}
The outdoor trials were conducted near the east entrance of ASU Bateman Physical Sciences Center H Wing, where the gNB and UE were separated by a large concrete wall. This configuration enabled a realistic evaluation of RIS-assisted mmWave coverage under mobility conditions. Similar to the indoor setup, the RIS was mounted at the gNB height and strategically positioned to maximize reflection coverage along the UE trajectory. The same measurement procedures were followed, with additional considerations such as reducing the gNB transmit power and aligning the gNB beam toward the RIS to mitigate environmental variability and multipath effects. The outdoor scene layout is illustrated in Fig.~\ref{outdoor-setups}.

\subsection{Experimental Link Validation} \label{sec:link-validation}
In this subsection, we describe the procedures used to collect and process data from the end-to-end system to validate the real-world performance of the proposed design. The data collection and processing steps are detailed below.
\textbf{Data Collection Procedure:}  
At each position along the UE trajectory (e.g., as shown in Fig.~\ref{fig:coverage_indoors}), a cloud-based application continuously monitored the xApp over a TCP link for valid UE RSRP and UE radio network temporary identifier(RNTI) values to determine the UE connectivity status. If the UE was not connected to the RAN, the RIS entered a sweeping mode, doing the exhaustive search while the xApp monitored for connectivity restoration.

Once the UE attached to the network—indicated by valid RSRP and RNTI values—a signal refinement procedure was triggered. This involved a local beam search around the last known RIS beam to identify the configuration that maximized RSRP. After identifying the optimal beam, the data collection module was triggered via the graphical user interface to log key metrics, including UE position, optimal RIS beam index, and corresponding performance data.

\textbf{Data Post-Processing:}  
The data samples collected in both the indoor and outdoor scenarios were post-processed to generate the results presented in Figs.~\ref{fig:coverage_indoors} and~\ref{fig:coverage_outdoors}. At each measurement point, joint RIS and UE beam sweeping was performed to construct a received power matrix (in dBm), along with the corresponding downlink throughput (in the indoor case). Using the optimal UE beam at each location, the RIS beam index with the highest received power was selected as the optimal RIS configuration. The throughput corresponding to this beam index was then extracted from the throughput matrix.

To estimate the background signal level without RIS assistance, the RIS was powered off and physically occluded. The RSRP was then recorded at selected positions and averaged to estimate the baseline signal level. Notably, in some locations, the UE could not connect to the gNB without RIS assistance. Furthermore, during the joint beam sweep, RSRP values corresponding to non-optimal RIS beams were often at or near the estimated noise floor. Consequently, in Figs.~\ref{fig:coverage_indoors} and~\ref{fig:coverage_outdoors}, the non-optimal RSRP and throughput values are presented as baseline measurements without RIS beamforming gain, while the optimal values represent measurements recorded with RIS beamforming applied.

\begin{figure}[h!]
    \centering
\subfigure[Indoor Scenario]{ \includegraphics[width=0.35\textwidth]{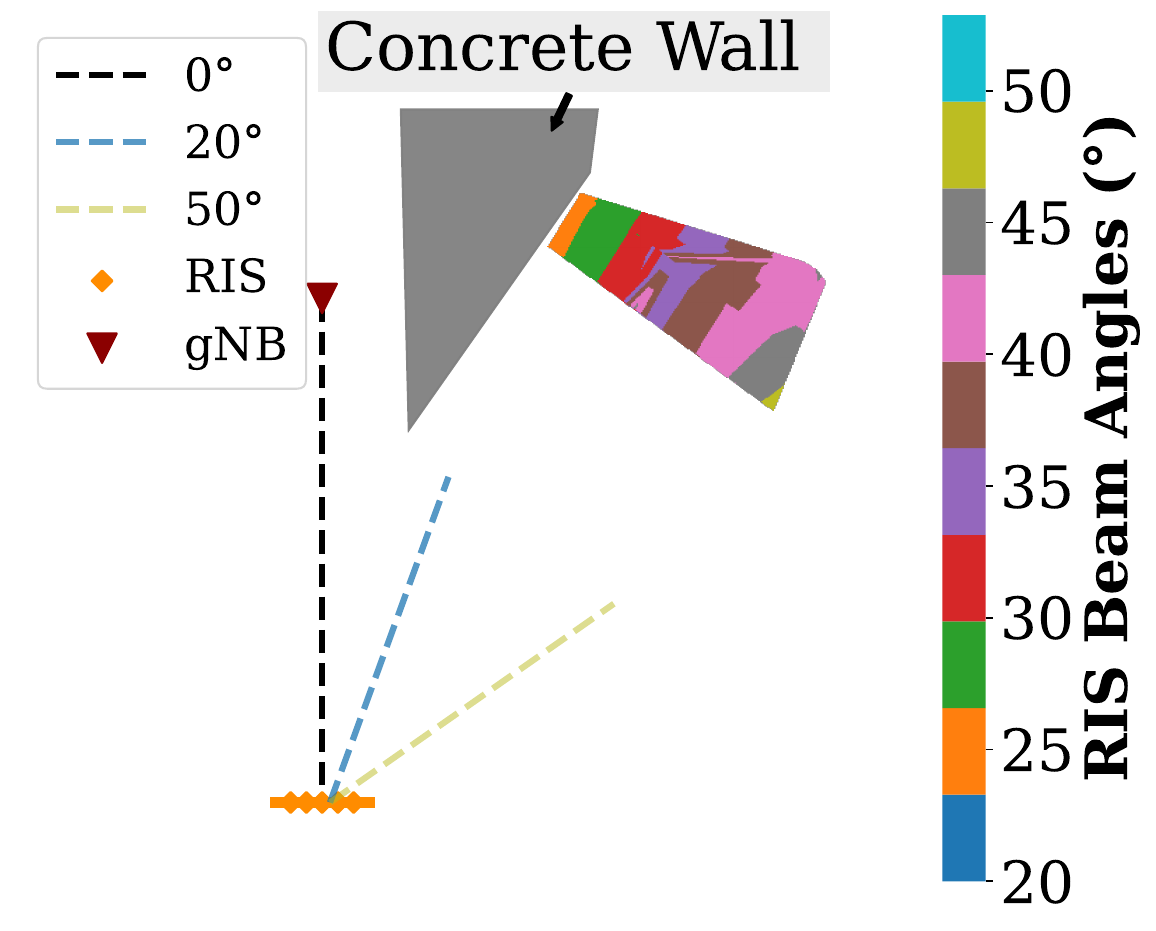}\label{indoor-beam:a}} 
\subfigure[ Outdoor Scenario]{ \includegraphics[width=0.35\textwidth]{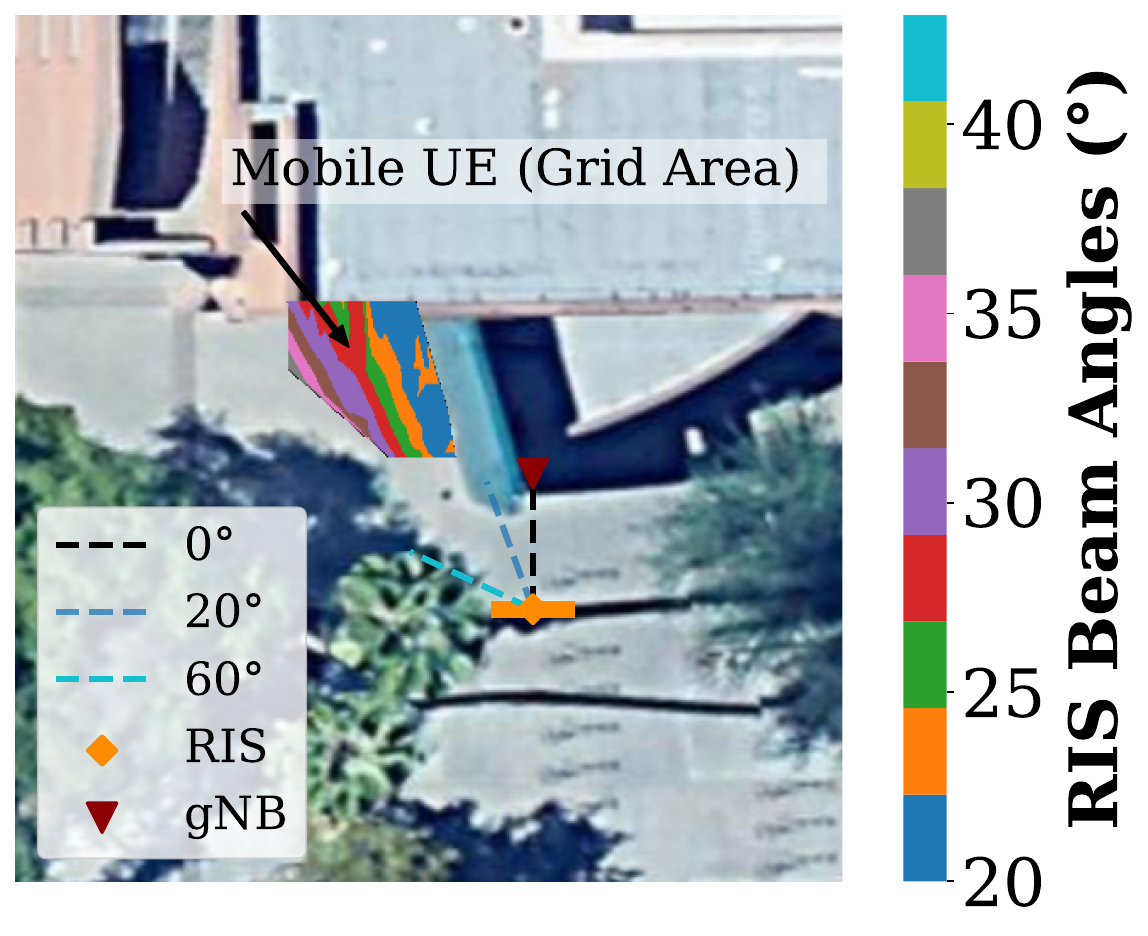} \label{outdoor-beam:b}}
   \caption{RIS beam distribution from indoor and outdoor field trials. (a) shows a mesh-grid heatmap of RIS beam angles corresponding to a subset of the predefined indoor UE trajectory. (b) presents a mesh-grid heatmap with linear interpolation based on outdoor measurements collected during mobile UE traversal.}

    \label{fig:beamcov} 
\end{figure}

\section{User Mobility Tracking}\label{sec:mobility}

RIS-assisted mmWave links hold significant potential for enhancing coverage in high-frequency cellular networks while also introducing unique challenges. RIS operates with narrow, reconfigurable beams, enabling UE to establish connections with the network at various static locations. However, ensuring seamless connectivity becomes increasingly challenging when the UE is in motion. Existing research~\cite{Sang24,Yang24} has primarily focused on evaluating coverage extension with fixed RIS beams at static locations. While effective in controlled scenarios, these approaches fail to address the dynamic nature of practical cellular environments, where UEs are rarely stationary. This limitation underscores the need for advanced beam management strategies to adapt to dynamic UE mobility.

To address this challenge, the integration of RIC platforms within the O-RAN architecture provides a promising solution for enabling dynamic beam control. By deploying xApps on the RIC, RIS beams can be adaptively adjusted in real-time to maintain robust communication links as the UE moves. In this paper, we propose and validate two approaches for managing UE mobility in RIS-assisted ORAN systems through field trials. The results demonstrate the effectiveness of these approaches in dynamically adapting RIS beams to ensure reliable connectivity in highly dynamic settings. In the subsections below, we present detailed descriptions of these methods and the test models adopted to evaluate the algorithms.

\subsection{Signal Power-Based Mobility Management}\label{sec:rsrp-based}

This work proposes mobility management algorithms that rely exclusively on signal strength, specifically utilizing reference signal received power (RSRP) values monitored by the xApp. These measurements are used to infer the mobility status of the UE and to inform decisions on dynamically updating the RIS beam configuration in response to observed RSRP trends. The following paragraphs present the two RIS-assisted mobility management approaches evaluated in this work, referred to as \textit{Algorithm~1} and \textit{Algorithm~2}.

\begin{figure}
    
    \centerline{\includegraphics[width=3in]{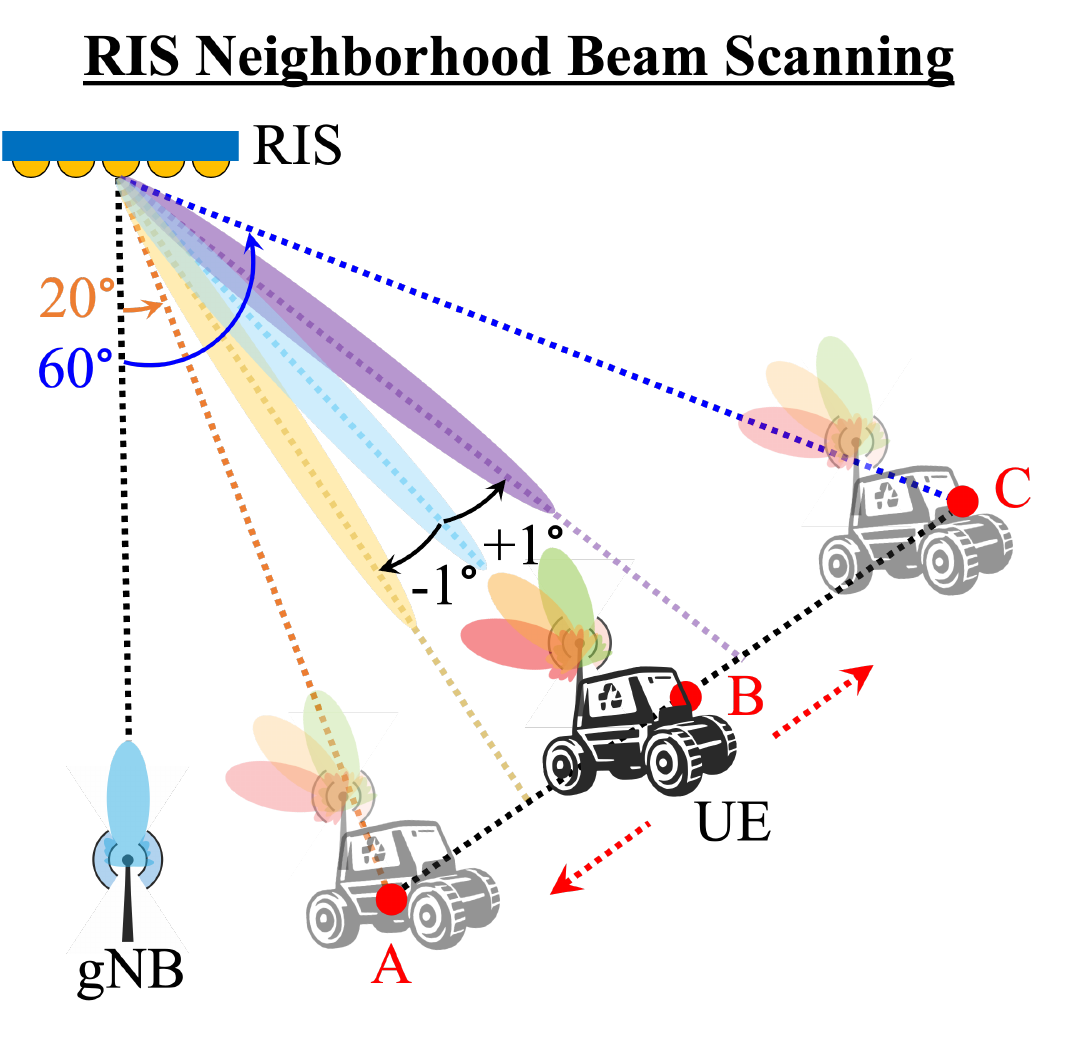}}
    
    \caption{Showcasing the continuous RIS neighborhood beam scanning approach to track the UE mobility.}
    
    \label{fig:ris_neighborhood_beam_scanning}

\end{figure}

\textbf{Algorithm~1: Continuous RIS Neighborhood Beam Scanning.} In this approach, the logic implemented in the xApp periodically toggles the RIS beam configuration between the current codeword and its two adjacent codewords in the beam codebook, specifically, the preceding and succeeding beams relative to the current selection (as shown in Fig.~\ref{fig:ris_neighborhood_beam_scanning}). After each probe, the corresponding RSRP is recorded, and the codeword that maximizes the instantaneous RSRP is selected for continued transmission. Due to the high-resolution beamsteering capability of the RIS, adjacent beams provides sufficient received power; thus, data transmission remains uninterrupted during the probing and switching process.

This approach enables smooth RIS beam transitions and maintains relatively steady signal power as the UE moves, since the candidate beams are typically close to the optimal RIS beam for the UE’s instantaneous location. While the method is simple and responsive to UE dynamics, it introduces moderate time overhead due to its continuous scanning nature. Furthermore, it may respond reactively to abrupt RSRP drops caused by misalignment between the RIS and UE beams or by dominant multipath components disrupting the optimal beam direction. These limitations motivate the design and evaluation of \textit{Algorithm~2}, as discussed in the following paragraph.

\textbf{Algorithm~2: RSRP Trend-Detector–Based RIS Beam Scanning.} This approach addresses the time overhead associated with Algorithm~1 by introducing a trigger-based mechanism for RIS beam updates. The xApp maintains a sliding window of the most recent $N$ RSRP samples, which are smoothed using a moving-average filter to suppress measurement noise. A non-parametric detector then classifies the resulting trend as either \emph{falling} or \emph{stable}. If a falling trend is detected, the RIS initiates localized neighborhood scanning, as described in Algorithm~1; otherwise, it retains the current beam configuration.

Compared to Algorithm~1, this method significantly reduces the time overhead associated with RIS beam reconfiguration. However, due to its  trigger-based nature, the selected RIS beams may lag behind or lead the UE’s optimal beam direction, particularly in cases where the UE beam remains fixed. These behaviors are evident in both the indoor and outdoor experiments, as shown in Fig.~\ref{exp_1:b} and Fig.~\ref{out-exp_2:b}, respectively.

\subsection{Algorithms Evaluation} \label{evaluation}

To evaluate the algorithms described in Section~\ref{sec:rsrp-based}, we utilized the setups and scenarios described in Section~\ref{sec:field}. The proposed mobility management algorithms were tested in both indoor and outdoor environments, as detailed below.
\subsubsection{Indoor Scenario}

In the indoor experiments, the UE was initially positioned at the edge of the RIS beam’s field of view. The establishment of the connection and the refinement of the signal strength were performed as described in Section~\ref{sec:link-validation}. Once a stable connection was achieved, the UE was moved along the predefined trajectories illustrated in Fig.~\ref{indoor-traj}. In \textit{Trajectory 1}, the UE traveled from the region corresponding to higher RIS beam angles toward the region with lower angles relative to the RIS boresight, that is, from point A to point C. In \textit{Trajectory 2}, the UE followed a more complex path: first moving from point A to point B, returning to point A, and then continuing to point C. The more dynamic mobility pattern in \textit{Trajectory 2} effectively demonstrates the capability of the proposed algorithms to maintain strong and reliable signal links under varying UE motion. The performance results of these tests are discussed in Section~\ref{sec:ue-move}. 

\subsubsection{Outdoor Scenario}

The outdoor scenario adopts similar test models to those used in the indoor experiments. As illustrated in Fig.~\ref{outdoor-traj}, \textit{Trajectory 1} involves the UE moving from point A to point C and then returning to point A. \textit{Trajectory 2} remains consistent with the indoor scenario, involving movement from point A to B, back to A, and then onward to C.

\begin{figure*}[h!]
	\centering
    \subfigure[EXP 1: Signal Strength]{ \includegraphics[width=0.458\textwidth]{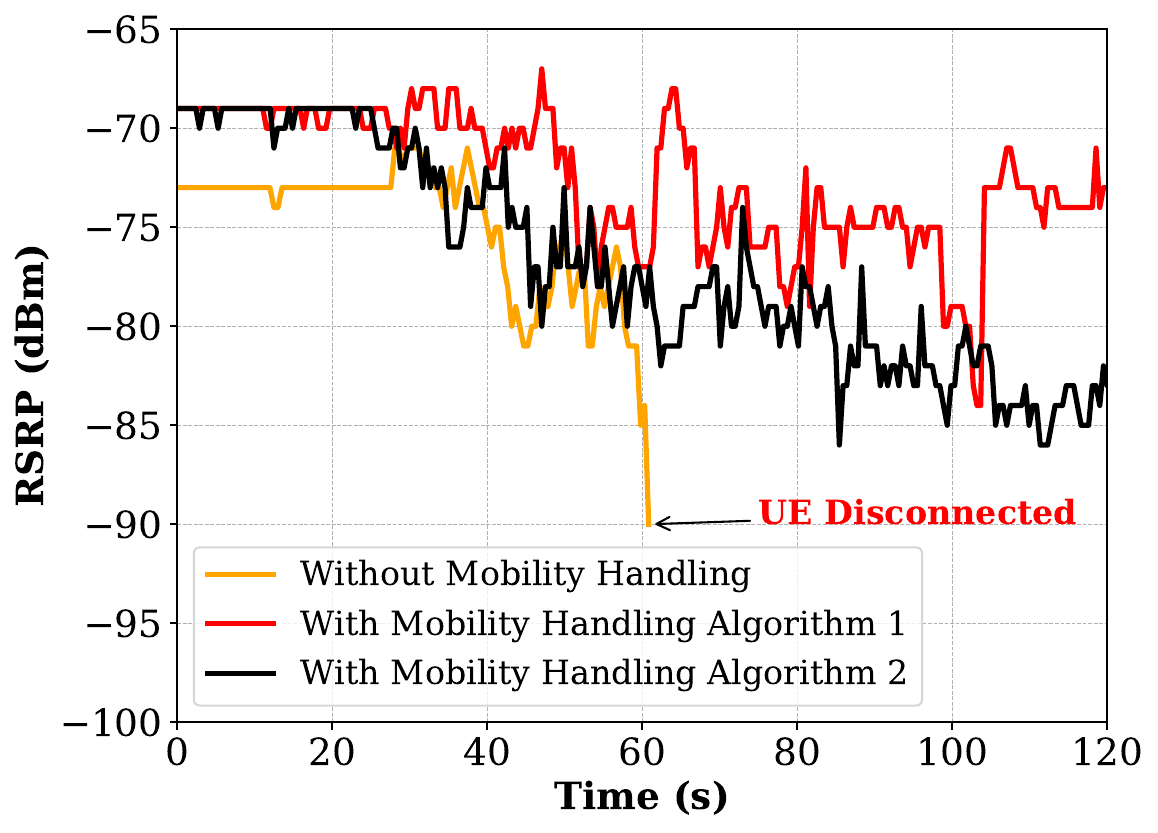} \label{exp_1:a} }
	\subfigure[EXP 1: Beam Angles]{ \includegraphics[width=0.48\textwidth]{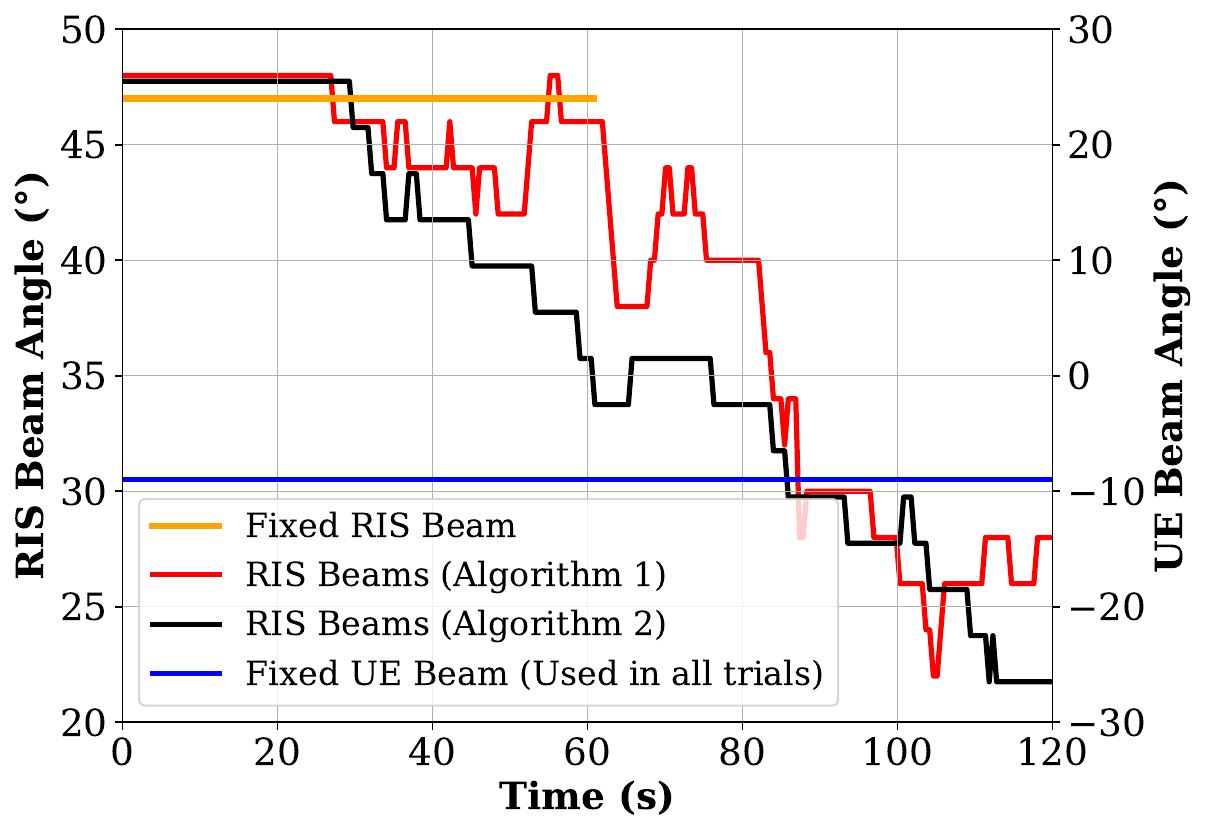} \label{exp_1:b} }

    \subfigure[EXP 2: Signal Strength]{ \includegraphics[width=0.458\textwidth]{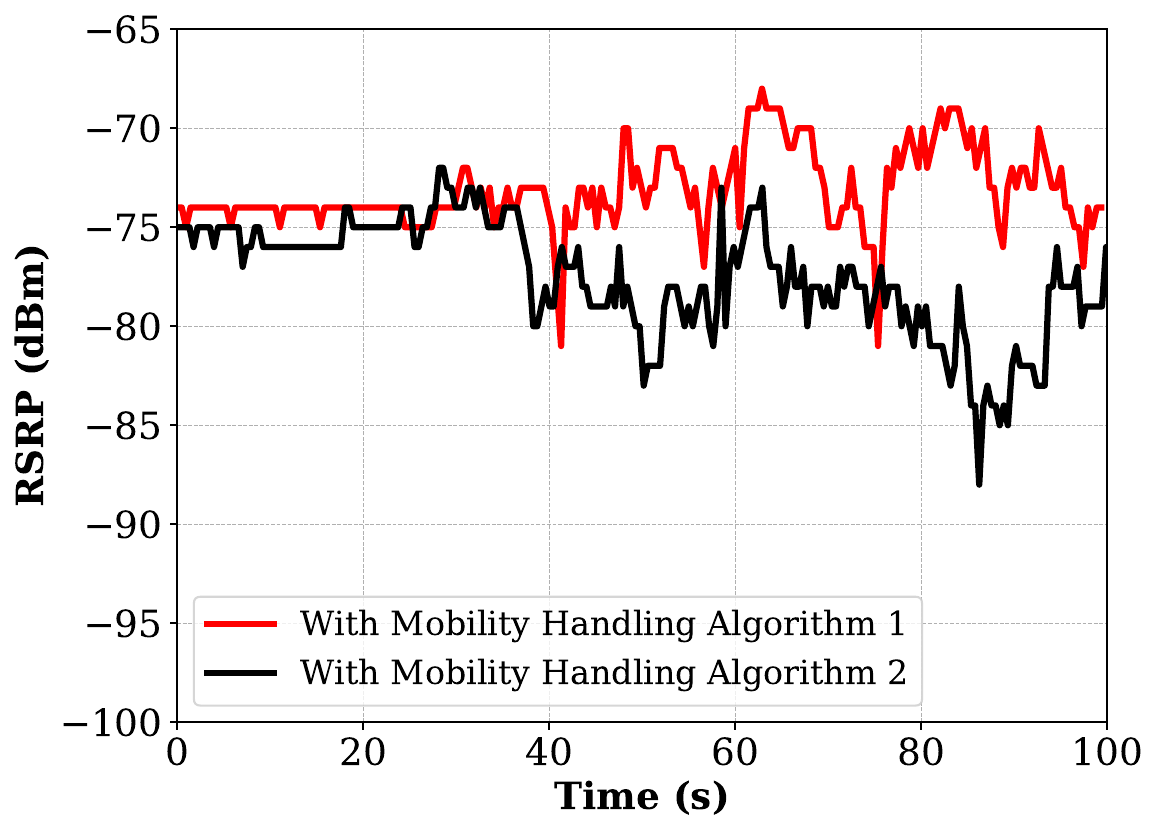} \label{exp_2:a} }
	\subfigure[EXP 2: Beam Angles]{ \includegraphics[width=0.48\textwidth]{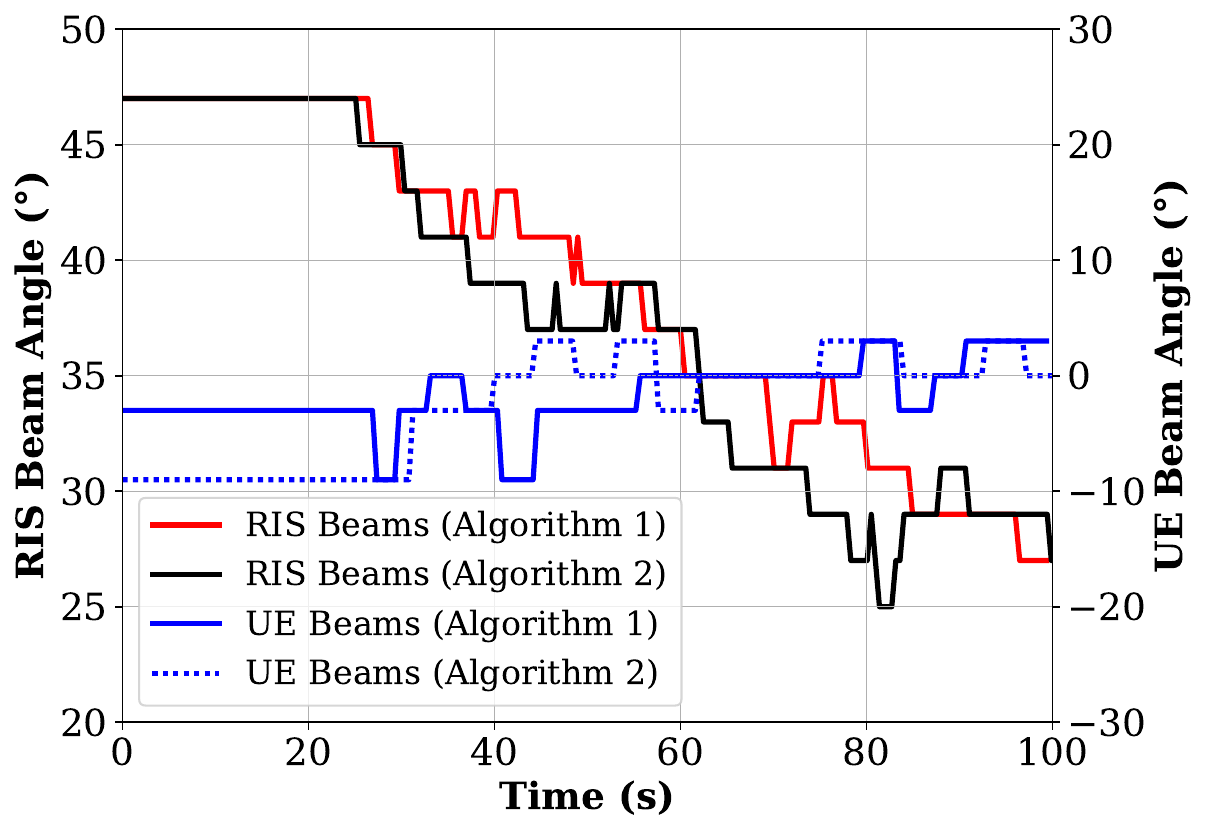} \label{exp_2:b} }

    \subfigure[EXP 3: Signal Strength]{ \includegraphics[width=0.458\textwidth]{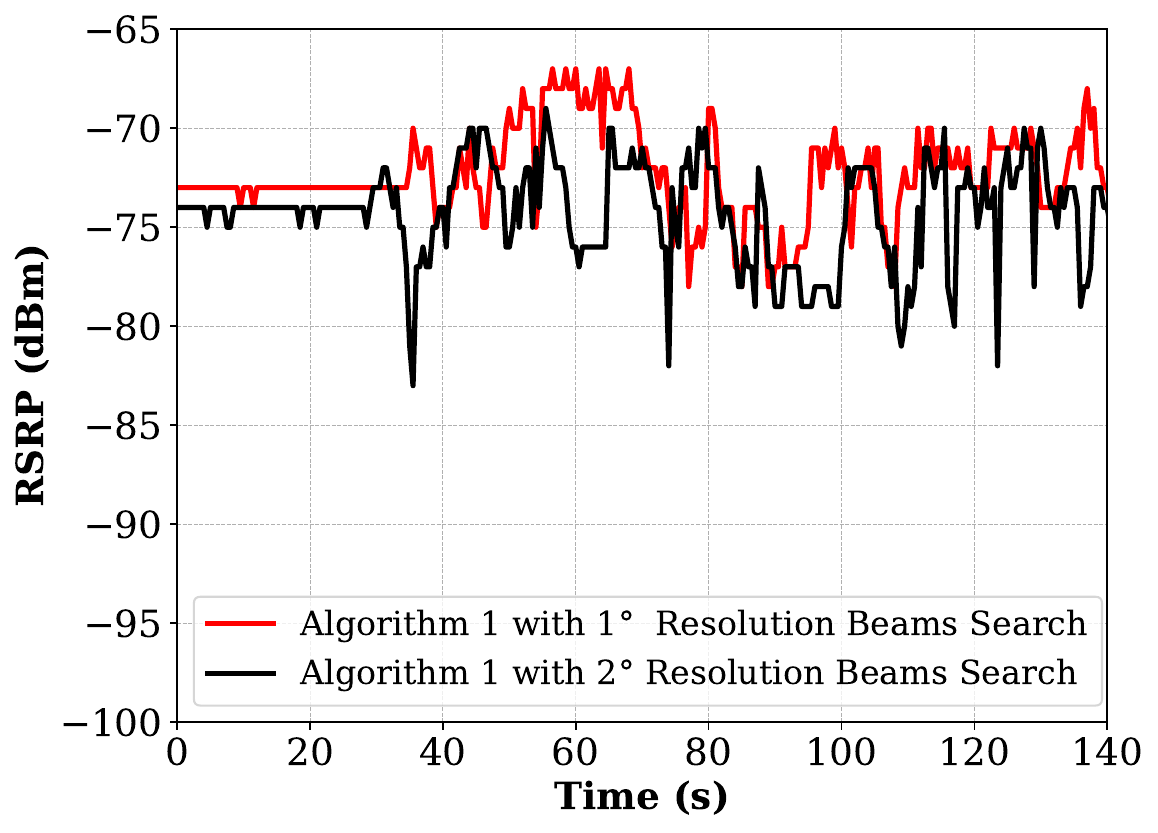} \label{exp_3:a} }
	\subfigure[EXP 3: Beam Angles]{ \includegraphics[width=0.48\textwidth]{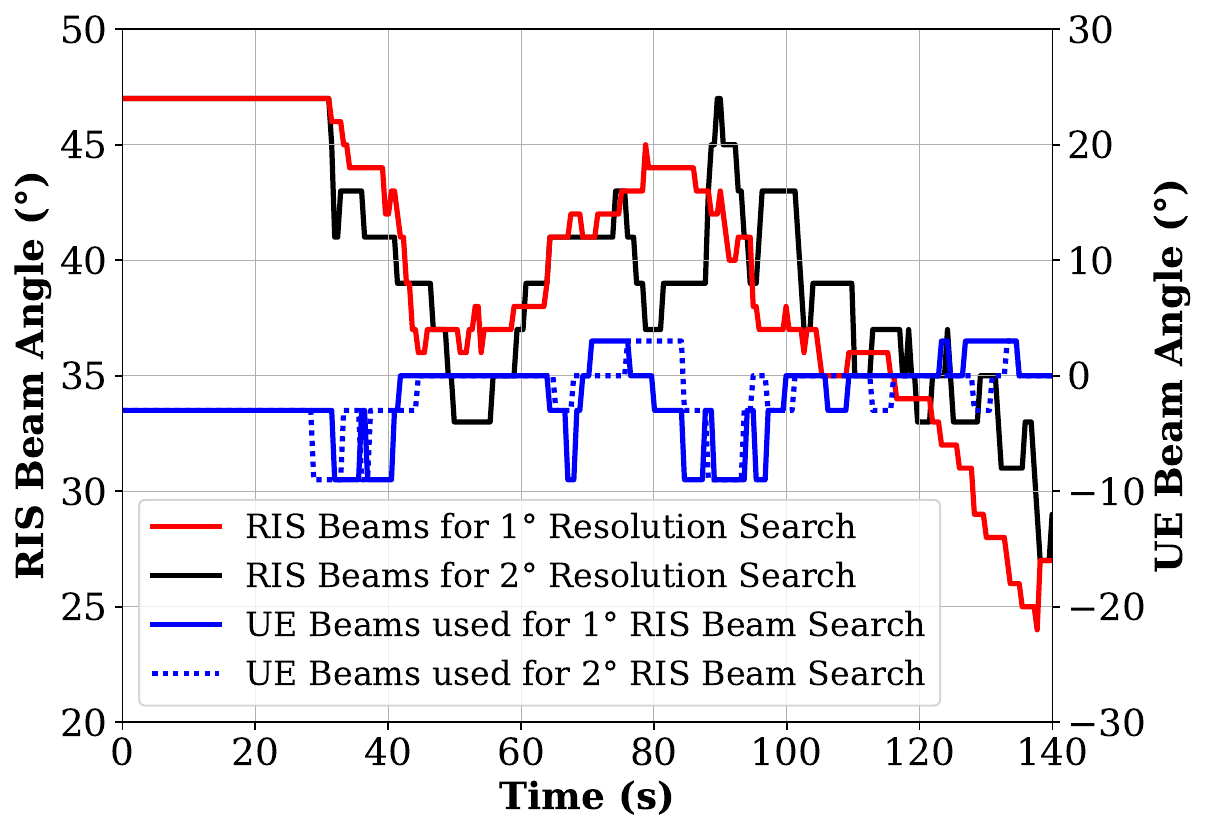} \label{exp_3:b} }

\caption{Indoor mobility test results for the movement trajectory depicted on the setup in Fig.~\ref{indoor-setups}. The first column presents real-time received power measurements, and the second column shows the corresponding real-time RIS and UE beam information.}

	\label{fig:mobility_test_indoor}
\end{figure*}

\section{Results and Analysis}\label{sec:results}

This section presents the results of field trials conducted to evaluate the coverage enhancement capabilities of the proposed mmWave RIS-assisted O-RAN communication system. In addition, it analyzes the performance of the proposed UE mobility management algorithms. The results are organized into two subsections: the first focuses on coverage measurements, while the second evaluates mobility management performance.

\subsection{Coverage Enhancement}\label{sec:coverage}

This subsection evaluates the improvements in signal strength and link throughput for the scenarios described in Section~\ref{sec:field}. The communication performance metrics monitored include the RSRP and the downlink (DL) throughput of the UE. These metrics are analyzed to quantify the performance gains achieved through RIS deployment and UE-side beam management.

\textbf{In the indoor scenario}, measurements were conducted along predefined UE trajectories under both RIS-enabled and baseline (no-RIS) conditions. The corresponding results are shown as scatter plots in Fig.~\ref{fig:coverage_indoors}, illustrating the RSRP and DL throughput, with color bars indicating values in dBm and Mbps, respectively. The deployment of the RIS resulted in substantial improvements in signal quality, with estimated RSRP gains ranging from 6~dB to 18~dB. As shown in Fig.~\ref{coverage:3b}, over 80\% of the collected samples exhibited an RSRP gain of approximately 10~dB, consistent with the indoor signal enhancement trends reported in~\cite{Sang24,Yang24}. 

Additionally, significant gains in downlink throughput were observed, with the average throughput increasing from 1236.31~Kbps to 2031.71~Kbps, corresponding to a system capacity improvement of approximately 64.3\%, as shown in Fig.~\ref{coverage:3b}.

\textbf{In the outdoor scenario}, the coverage evaluation focused primarily on signal power enhancement. As illustrated by the heatmaps in Fig.~\ref{fig:coverage_outdoors}, the received signal power improved significantly with RIS deployment. RSRP gains across the UE trajectory grid points reached up to 19~dB, with an average gain of approximately 15~dB. Furthermore, we evaluated RIS beam configurations designed to achieve broader field-of-view coverage and finer angular resolution. As detailed in Section~\ref{sec: RIS DESIGN}, the deployed RIS system employs codewords that span a wide range of reflection angles, thereby enhancing coverage performance in dynamic user scenarios.

Figure~\ref{fig:beamcov} presents linear heatmaps of the optimal RIS beam configurations corresponding to various UE positions in both indoor and outdoor scenarios. For the outdoor case, 70 data points were collected across five rows using joint RIS and UE beam sweeping. In each instance, the optimal RIS beam was selected based on the UE beam that yielded the highest received signal strength. The indoor heatmap was generated from a subset of the trajectory samples shown in Fig.~\ref{fig:coverage_indoors}. The color distributions across the mesh grids in both subplots of Fig.~\ref{fig:beamcov} demonstrate that the designed RIS codewords effectively steer reflected signals toward the intended UE positions.

These results highlight the capability of the RIS system to dynamically adapt to varying UE positions, thereby mitigating coverage gaps in previously underserved or weak-signal regions. Moreover, the successive beam transitions, captured by the smooth color variations across Fig.~\ref{fig:beamcov}, support seamless UE mobility management by maintaining consistent beam alignment throughout the trajectory. Overall, the findings underscore the potential of RIS technology to enhance both coverage flexibility and reliability in mmWave communication systems.

\begin{figure*}[h!]
	\centering
    \subfigure[EXP 1: Signal Strength]{ \includegraphics[width=0.435\textwidth]{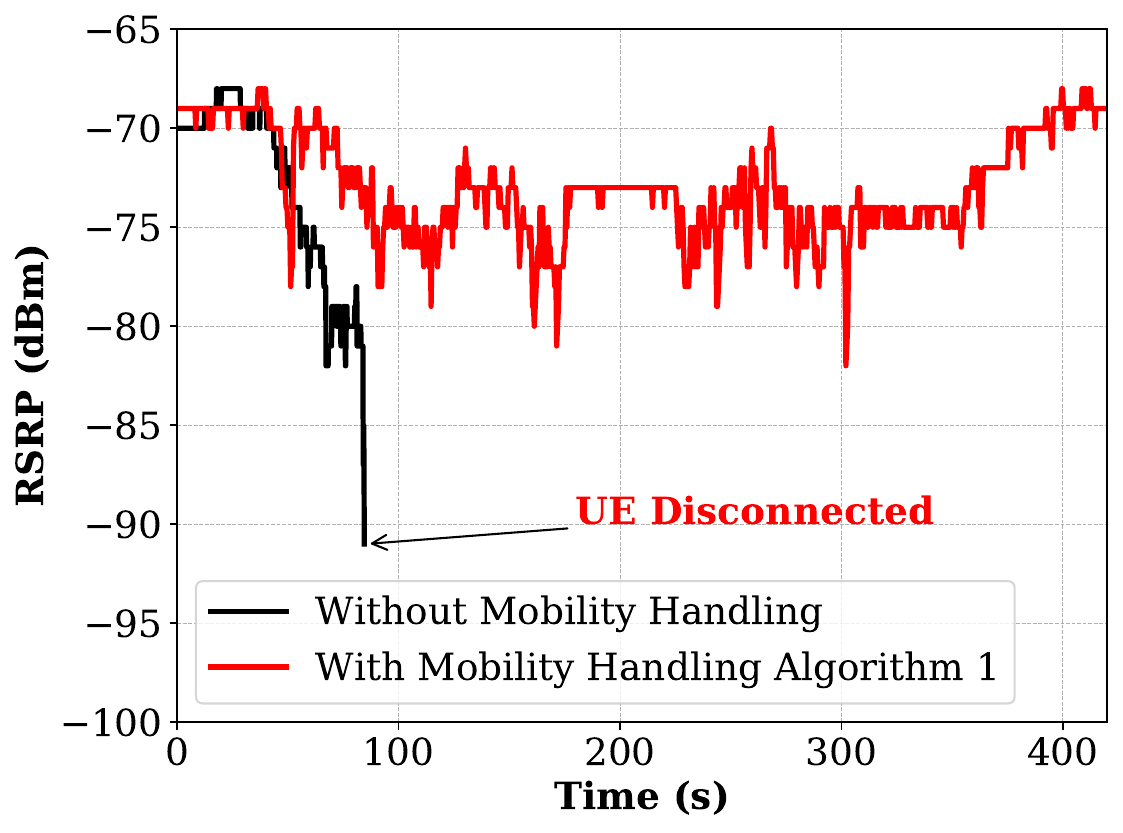} \label{out-exp_1:a} }
	\subfigure[EXP 1: Beam Angles]{ \includegraphics[width=0.465\textwidth]{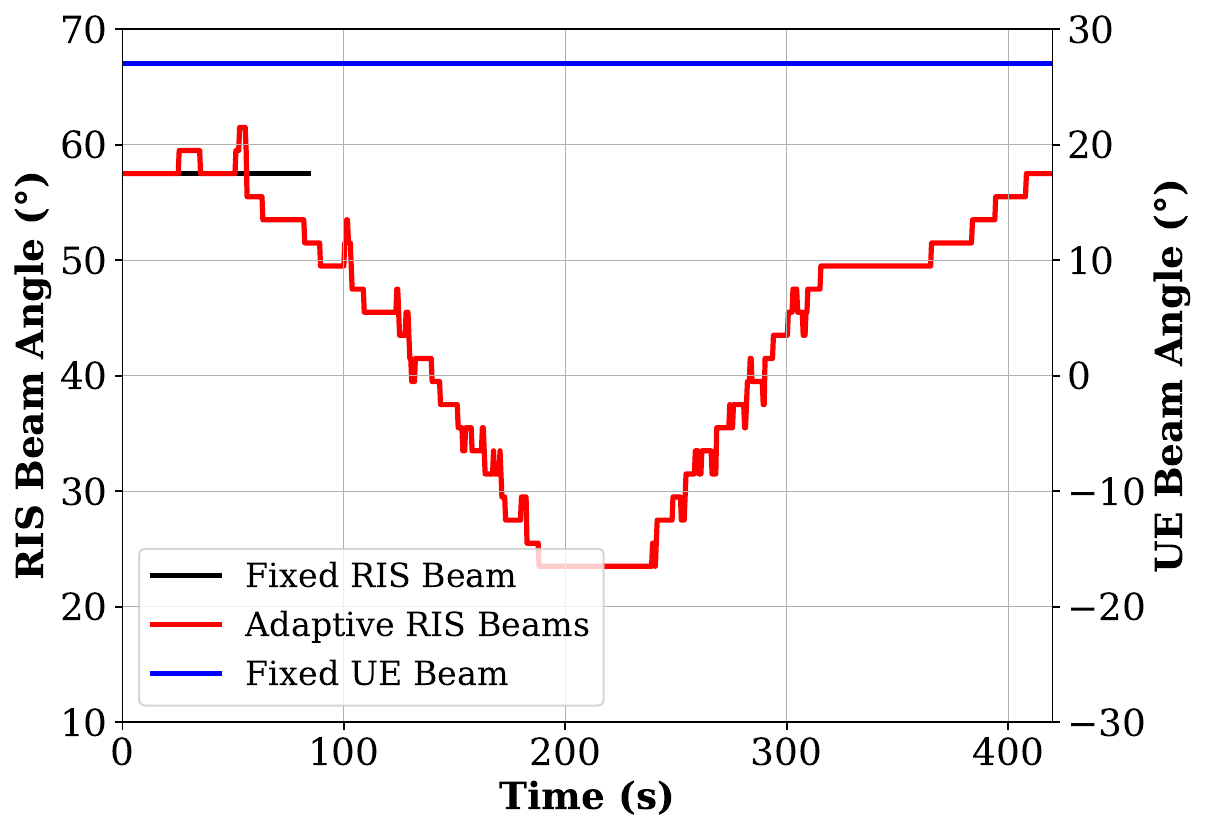} \label{out-exp_1:b} }

    \subfigure[EXP 2: Signal Strength]{ \includegraphics[width=0.435\textwidth]{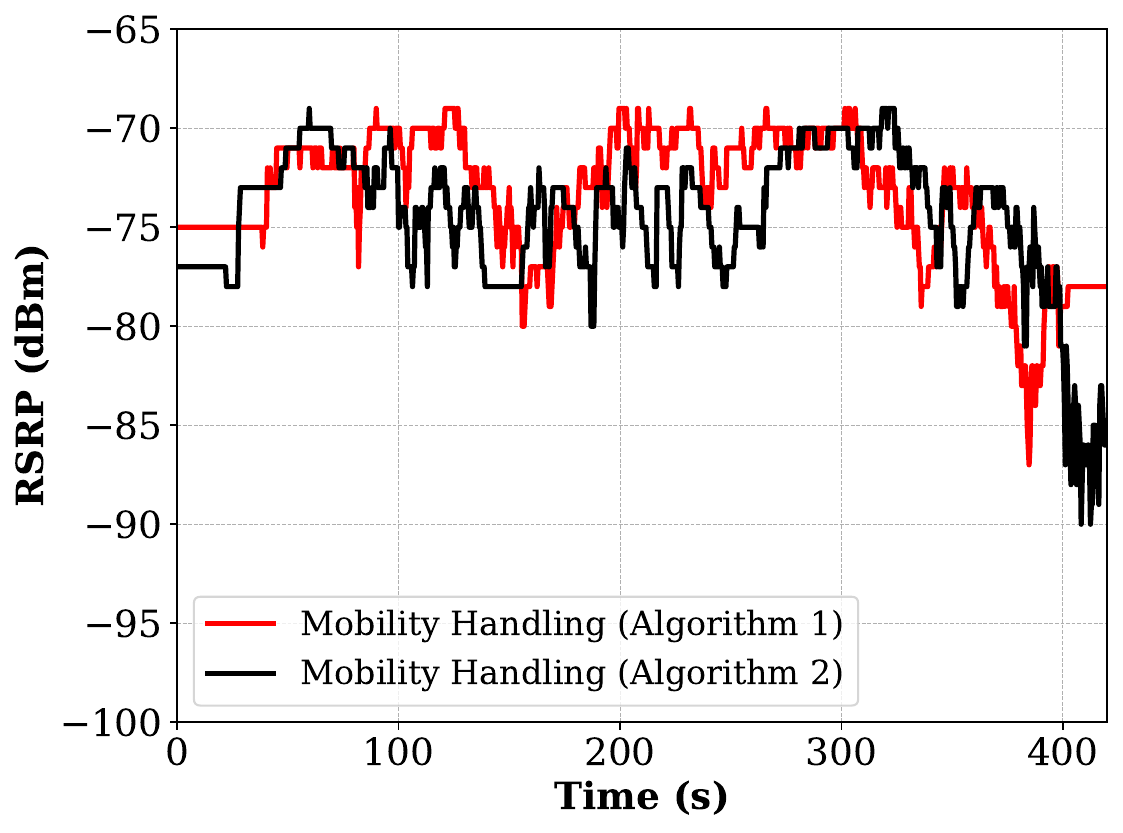} \label{out-exp_2:a} }
	\subfigure[EXP 2: Beam Angles]{ \includegraphics[width=0.465\textwidth]{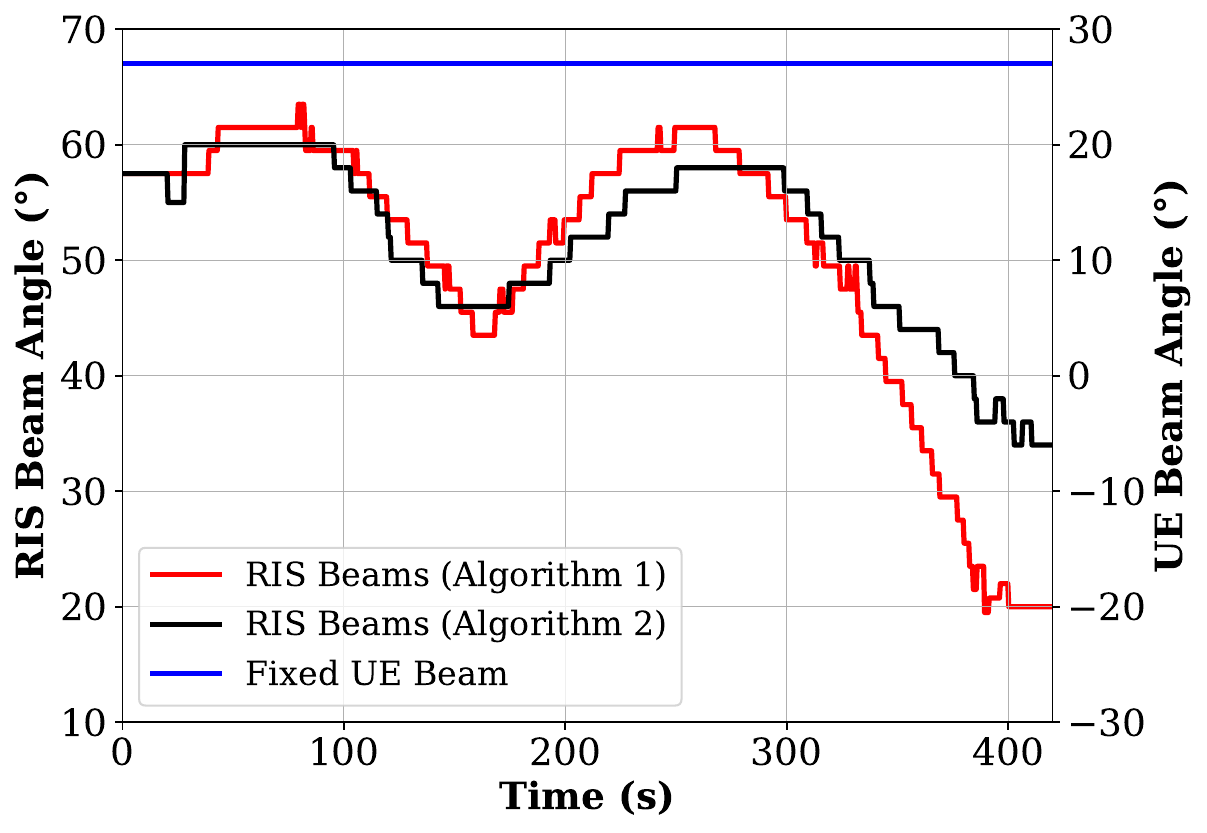} \label{out-exp_2:b} }

   \subfigure[EXP 3: Signal Strength]{ \includegraphics[width=0.445\textwidth]{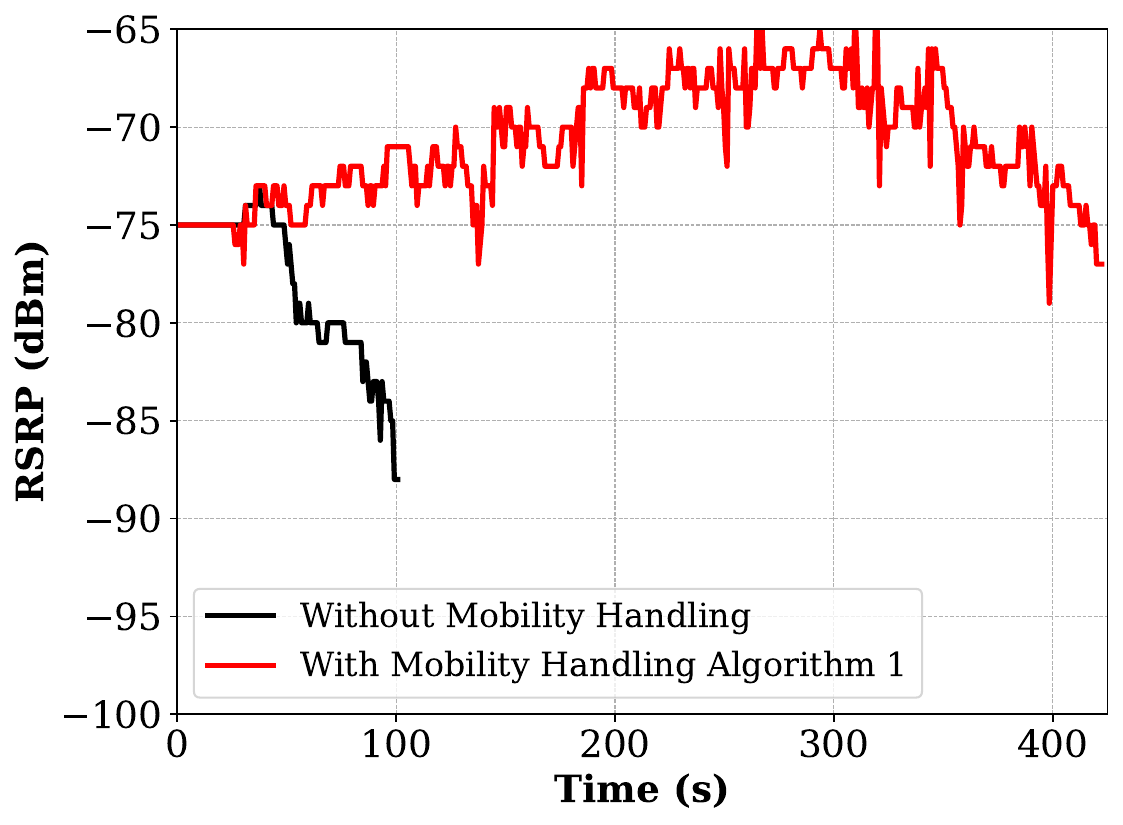} \label{out-exp_3:a} }
	\subfigure[EXP 3:  Beam Angles]{ \includegraphics[width=0.450\textwidth]{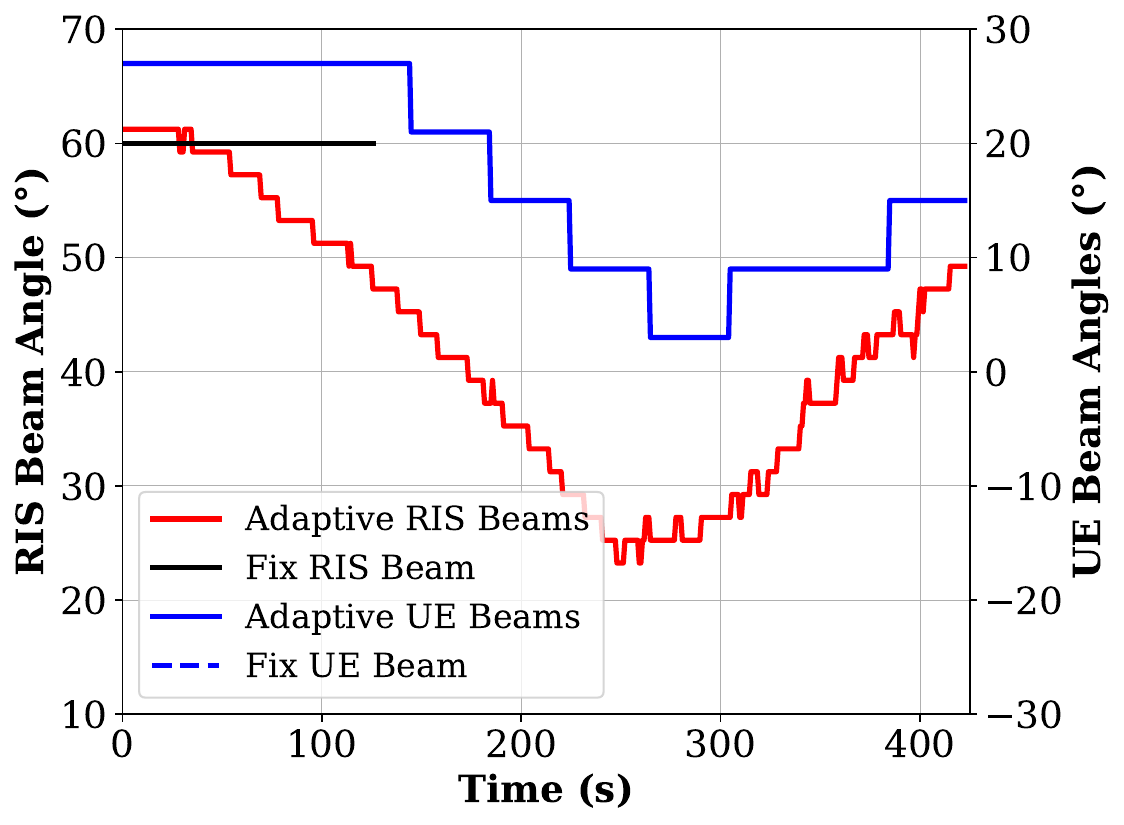} \label{out-exp_3:b} }

\caption{Mobility test results for the experiments conducted outdoors. The first column presents the received signal strength trends recorded in real-time from the O-RAN platform during each experiment. The second column illustrates the evolution of the RIS and UE beam indices throughout the experiments.}

	\label{fig:mobility_test_outdoors}
\end{figure*}

\subsection{UE Mobility Management}\label{sec:ue-move}

In this subsection, we present the results of RIS-assisted UE mobility handling under two experimental settings. The objective is to demonstrate the effectiveness of the proposed RSRP-based mobility management algorithms in supporting a UE moving at a speed of 5~cm/s. We compare RSRP trends across different configurations: with and without mobility management, and with fixed versus adaptive UE beam configurations.

\subsubsection*{\textbf{Indoor Trials}}
The next paragraphs present and analyze the results from three distinct indoor experiments, each designed to evaluate different aspects of RIS-assisted UE mobility under controlled conditions.

\paragraph{Impact of Adaptive RIS Beams}
The results shown in Fig.~\ref{fig:mobility_test_indoor} (EXP~1) illustrate signal strength dynamics and RIS beam configurations during UE mobility, comparing scenarios with and without RIS beam tracking following the initial link establishment. The RSRP trends recorded during \textit{Trajectory~1} indicate that the UE experiences disconnection after moving a certain distance from its starting position. This highlights the challenge of maintaining stable connectivity in RIS-assisted mmWave links without dynamic beam adaptation.

As shown in Fig.~\ref{exp_1:a}, a gradual decay in RSRP is observed over time, even when the RIS beam is optimally updated. This decline results from the use of a fixed UE beam: although the initial alignment between the RIS and UE beams is optimal, UE movement introduces angular misalignment, leading to signal degradation. This limitation motivated the experiments in Figs.~\ref{exp_2:a} and~\ref{exp_2:b} (EXP~2), in which the UE beam is adaptively updated in parallel with the RIS beam. This joint adaptation smooths RIS beam transitions under Algorithm~1 and results in more stable RSRP levels compared to the fixed UE beam scenario. The difference in initial RSRP levels between Figs.~\ref{exp_1:a} and~\ref{exp_2:a} is attributed to variations in the initial RIS beams selected by the beam-sweeping algorithm, as well as differing experimental conditions across trials. As observed in Fig.~\ref{exp_1:a}, the red, black, and orange curves exhibit RSRP variation due to different initial beam selections.

\paragraph{Comparison of the Two Algorithms}
Algorithm~2 is more resource-efficient, as it triggers RIS beam searches only when necessary. However, it requires careful tuning of parameters such as the RSRP window size and the threshold for detecting RSRP drops. These parameters are site-specific and must be empirically optimized for both indoor and outdoor deployments.

In contrast, Algorithm~1 performs continuous RIS beam searches, improving its ability to recover from suboptimal beam configurations. While this enhances robustness in fast-varying environments, it increases both computational and signaling overhead. Moreover, incorrect RIS beam selections can propagate across iterations, resulting in misalignment between the predicted and actual UE positions.

This behavior is illustrated in Fig.~\ref{exp_1:b}, where, during the initial test phase (30–60~s), Algorithm~1 attempts to recover from a suboptimal RIS beam configuration, as evidenced by the RSRP spikes observed in Fig.~\ref{exp_1:a}. In comparison, Algorithm~2 exhibits slower but more stable recovery behavior, leading to lower average RSRP levels—highlighted by the differences between the red and black curves.

The RSRP and RIS beam dynamics observed with Algorithm~2 in Figs.~\ref{exp_1:a} and~\ref{exp_1:b} align with expectations for a reactive, threshold-based approach. In contrast, Algorithm~1’s sensitivity to instantaneous RSRP values makes it more susceptible to multipath effects and background interference, resulting in less stable beam adaptation.

\paragraph{Impact of Finer Resolution RIS Beam Search}
In EXP~3, we evaluated the proposed mobility algorithm under different RIS beam search resolutions. A $1^\circ$ resolution implies beam searches among neighbors separated by $1^\circ$. As shown in Figs.~\ref{exp_3:a} and~\ref{exp_3:b}, RIS beam transitions are smoother with $1^\circ$ resolution compared to $2^\circ$. Additionally, the RSRP values in Fig.~\ref{exp_3:a} exhibit reduced fluctuation when using the finer-resolution beam codebook. These results suggest that, with faster xApp update rates, higher-resolution RIS codebooks, and sufficient search resources, it is possible to maintain a highly stable RIS-assisted mmWave link with consistent signal quality.

\subsubsection*{\textbf{Outdoor Trials}}

We conducted two outdoor experiments to evaluate the impact of different UE mobility patterns and RIS beam-tracking algorithms, as illustrated in Fig.~\ref{fig:mobility_test_outdoors}. The experiments are organized as follows: EXP~1 examines the effect of RIS beam tracking in outdoor environments, while EXP~2 compares the performance of the two proposed beam-switching algorithms. Both experiments were performed with a fixed UE beam configuration.

In the first experiment (EXP~1), the goal was to evaluate the proposed mobility algorithm in a dynamic outdoor environment. The UE moved across the RIS field of view, starting at approximately 60$\degree$, traversing to 20$\degree$, and returning, representing a path from point A to C through B, as shown in Fig.~\ref{outdoor-traj}. With a fixed UE beam, the RSRP trend in Fig.~\ref{out-exp_1:a} shows expected decay and rise as the UE traces the trajectory. The xApp was configured to report RSRP every 50~ms. The RSRP trend was smoother with a 2$\degree$ nearest-neighbor RIS beam search due to reduced multipath interference compared to the indoor scenario. The RSRP and RIS beam angle updates were compared with and without the mobility management algorithm. Without the algorithm, the UE experienced connectivity loss. In contrast, with the algorithm enabled, the connection was successfully maintained, demonstrating the effectiveness of mobility-aware RIS adaptation in outdoor environments.

In the second experiment(EXP~2), we compared the two algorithms under a more dynamic mobility trajectory. The UE moved from 60$\degree$ to 45$\degree$, returned to 60$\degree$, and then proceeded to 20$\degree$. This trajectory involved multiple beam transitions, allowing for comprehensive evaluation of tracking performance. The xApp RSRP reporting interval was set to 100~ms. The results show that Algorithm~2, due to its event-based design, incurs lower computational overhead. However, this efficiency comes at the cost of reduced tracking accuracy. For example, it incorrectly identified the optimal RIS beam as 36$\degree$ instead of the true 20$\degree$, leading to degraded link performance, especially around the 400th seconds in Fig.~\ref{out-exp_2:a}, where a significant RSRP drop is observed

The third experiment (EXP~3) highlights the significance of jointly adapting the RIS and UE beams using Algorithm~1. In this experiment, the logic of Algorithm~1 is applied to both the RIS and the UE beams. However, the neighboring UE beams are only  occasionally searched and updated because the UE has significantly wider beamwidth as compared to RIS. The UE follows a trajectory similar to that in EXP~1. With the adaptive UE beams, the signal strength observed in Fig.~\ref{out-exp_3:a} remains high and steady, in contrast to the fluctuating signal strength shown in Fig.~\ref{out-exp_1:a} with a fixed UE beam. This result indicates that, with incorporating the UE beam sweeps, the RSRP trend becomes much more stable and the probability of UE disconnection is drastically reduced.

\subsection{Discussion}

The experimental results clearly demonstrate that effective UE mobility management in RIS-assisted mmWave systems requires coordinated adaptation of both RIS and UE beam configurations. Algorithm~1 provides robust tracking through its continuous search mechanism but incurs higher computational overhead and is more sensitive to environmental reflections. In contrast, Algorithm~2 offers greater resource efficiency and smoother performance trends, though it may exhibit delayed response during rapid mobility transitions due to its reactive nature. The results also highlight that UE beam misalignment can significantly limit the performance gains achievable with RIS, in both indoor and outdoor settings. Incorporating adaptive UE beam tracking notably improved signal consistency and reduced beam-switching noise. Outdoor evaluations confirmed the repeatability of both algorithms, while also revealing their limitations in handling abrupt direction changes. These findings motivate future research on enhanced UE codebooks, hybrid learning-based prediction schemes for proactive beam switching, and closed-loop optimization strategies to support high-mobility scenarios.

\section{Conclusion} \label{sec:conc}

In this paper, we leveraged the existing O-RAN standard to implement and validate a RIS-aided mmWave link. The RIS prototype used in this work consists of 1024 elements arranged in a \(32 \times 32\) configuration, with each unit cell featuring 1-bit phase resolution and independent programmability. Comprehensive field trials were conducted in indoor settings to evaluate the practical performance of RIS for coverage enhancement within a private 5G testbed operating in the FR2 band. The experimental results demonstrate notable improvements in communication quality, including enhanced signal strength and stable connectivity in challenging environments. Additionally, a UE mobility management algorithm was developed to maintain connectivity under dynamic conditions by adaptively adjusting the RIS beams based on sequential patterns in the communication metrics obtained via the O-RAN interfaces. This work establishes a foundation for integrating our RIS system into O-RAN 5G networks, demonstrating that RIS can extend coverage and enable seamless connectivity for mobile users. Future work will: (i) scale the 5G testbed to wider bandwidths and larger subcarrier spacings, (ii) optimize RIS configurations for multi-user operation, and (iii) incorporate AI/ML-based prediction for joint RIS and UE beam management under dynamic conditions.

\balance

\end{document}